\documentclass[12pt,a4paper]{article}%

\usepackage{amsmath,amssymb,amsfonts}
\usepackage{caption2}
\usepackage[]{hyperref}
\usepackage[pdftex]{color,graphicx}
\usepackage{multicol}

\numberwithin{equation}{section} 
\setcaptionmargin{1cm}
\newcommand{\hhref}[1]{\href{http://arxiv.org/abs/#1}{{\it arXiv:#1}}}
\newcommand{\ba}{\begin{eqnarray}}
\newcommand{\ea}{\end{eqnarray}}
\newcommand{\no}{\nonumber}

\begin{document}

\begin{titlepage}
\begin{flushright}
LA-UR-12-21211\hfill
\end{flushright}
$\quad$
\vskip 2.0cm
\begin{center}
{\huge \bf  On Partial Compositeness and the\\ CP Asymmetry in Charm Decays} 
\vskip 1.0cm {\large {\bf Boaz Keren-Zur}$^a$, {\bf Paolo Lodone}$^a$, {\bf Marco Nardecchia}$^b$, {\bf Duccio~Pappadopulo}$^a$, {\bf Riccardo~Rattazzi}$^a$, and {\bf Luca Vecchi}$^c$} \\[1cm]
{\it 
$^a$ Institut de Th\'eorie des Ph\'enom\`enes Physiques, EPFL, Lausanne, Switzerland \\
$^b$ CP3-Origins \& 
DIAS, University of Southern Denmark, 
Odense M, Denmark \\
$^c$ Theoretical Division T-2, MS B285, Los Alamos National Laboratory, Los Alamos, 
USA
} \\[5mm]
\vskip 1.0cm
\today
\end{center}

\begin{abstract}


Recently, the LHCb and CDF collaborations reported the measure of an unexpectedly large direct CP asymmetry in D meson decays. 
In this paper we ask if new physics associated with Partial Compositeness could plausibly explain this result.
We find that Composite Higgs models with mass scale around 10 TeV can account for it, while marginally satisfying all other flavor constraints in the quark sector. The minimal framework is however inadequate in the lepton sector due to the strong constraint from $\mu \rightarrow e \gamma$. This tension can be efficiently alleviated by realizing Partial Compositeness within Supersymmetry. The resulting models can saturate the CP asymmetry in D decays for superpartner masses close to the TeV scale and somewhat large A-terms. The supersymmetric realization of Partial Compositeness also offers a predictive and phenomenologically viable organizing principle for R-Parity violation, and may result in very distinctive signatures at hadron colliders.
With or without Supersymmetry, the neutron EDM is expected to be around the present experimental sensitivity. 

\end{abstract}
\end{titlepage}


\section{Introduction}

After a year of running, the Large Hadron Collider (LHC) has already started exploring physics at the weak scale. Although no evidence of physics beyond the standard model (SM) has been found so far, two interesting pieces of data emerge from the 2011 run. One is consistently observed by both ATLAS and CMS, and might in fact be the first indication of the SM Higgs boson~\cite{Higgs}. The other is the measurement for the first-time of direct CP-violation in D meson decays, originally reported by LHCb~\cite{Aaij:2011in} and subsequently confirmed by the Tevatron CDF~\cite{CDFupdate}.

While the existence of a SM Higgs boson not far above the LEP bound was anticipated by numerous indirect measurements, the observation of large CP violation in the charm sector appeared somewhat as a surprise. Defining $a_f$ to be the time-integrated CP asymmetry for the process $D^0\to f$, and combining the LHCb and CDF results assuming Gaussian and fully uncorrelated uncertainties, one finds~\cite{CDFupdate}:
\begin{equation} \label{eq:deltaAcp}
\Delta a_{CP} = a_{K^+K^-}-a_{\pi^+\pi^-} = (- 0.67 \pm 0.16) \%.
\end{equation}
This measurement deviates by $\sim3.8\, \sigma$ from the no-CP-violation hypothesis, and represents the first evidence of direct CP violation in D meson decays. Perhaps even more strikingly, though, is the fact that such value is significantly larger than the naive SM estimate $\Delta a_{CP} = O(0.1)\%$, see e.g. \cite{Cheng:2012xb}. 

The magnitude of the irreducible hadronic uncertainties involved in the determination of the SM prediction makes it impossible at the moment to establish whether the above result can be accounted for entirely by the SM, or if~(\ref{eq:deltaAcp}) should actually be seen as indication of new physics. This ambiguity stimulated a number of theoretical efforts aiming to accommodate the result within the SM~\cite{Golden:1989qx}-\cite{Brod:2012ud} or new physics scenarios~\cite{Hinchliffe:1995hz}-\cite{Cheng:2012xb}, see also \cite{Isidori:2012boh}. 

Assuming that the measurement~(\ref{eq:deltaAcp}) is due to physics beyond the SM, then a widely held opinion is that this new physics must possess a highly non-trivial flavor structure given the strong constraints from all the other flavor observables. An attractive possibility is that the very same organizing principle responsible for generating the SM flavor hierarchy also controls flavor-violation in the new physics sector. The goal of the present paper is to investigate under which conditions and to which extent Partial Compositeness can be such a principle.

Partial Compositeness is a seesaw-like mechanism that naturally explains the hierarchy among the SM fermion masses. It was first discussed in~\cite{Kaplan:1991dc} as an alternative way to address the flavor problem in Technicolor models, and subsequently realized thanks to the gauge/gravity correspondence via wavefunction localization along the extra-dimension in Randall-Sundrum models, see e.g.~\cite{Gherghetta:2000qt}\cite{Agashe:2004cp}. Most of the literature on the subject realizes the Partial Compositeness paradigm within composite Higgs models~\cite{CH} (see also~\cite{Contino:2006nn} for a recent discussion), whereas the analysis of the phenomenological implications of Supersymmetric (SUSY) realizations was initiated in~\cite{flavorful1}. 

We will review Partial Compositeness and its solution of the SM flavor puzzle in Section~\ref{sec:general}. The conditions under which Composite Higgs models implementing this mechanism can explain~(\ref{eq:deltaAcp}) will be critically analyzed in Section \ref{sec:comp}, whereas SUSY realizations will be discussed in Section \ref{sec:susy}, where the results of~\cite{flavorful1} will be generalized to include scenarios with R-parity violation. We will finally present our conclusions in Section~\ref{sec:conclusions}.


\section{Partial Compositeness}  \label{sec:general}
\label{framework}

Let us start by briefly reviewing the paradigm of Partial Compositeness, and how this idea can be used to explain the SM flavor hierarchy. 
The basic assumption is that at the UV cutoff $\Lambda_{\rm UV}$ the SM fermions $f_i^a$ 
couple linearly to operators ${\cal O}_i^a$ of a confining, flavorful sector:
\ba\label{linear}
\lambda_i^a f_i^a{\cal O}_i^a \,\, ,
\ea
where hereafter $a,b,\dots=q,u,d,\ell,e$ and $i,j,\dots=1,2,3$ denote flavor and family indices, respectively. In addition to~(\ref{linear}) the two sectors communicate via the weak gauging of the SM group, taken to be a subgroup of the chiral symmetry of the new dynamics. 


Using naive dimensional analysis (NDA), and adopting the notation of \cite{Giudice:2007fh}, one finds that the low energy effective Lagrangian renormalized at the confinement scale $m_\rho$ of the flavor sector schematically reads:
\begin{equation}\label{NDA}
\mathcal{L}_{\rm NDA} = \frac{m_{\rho}^4}{g_{\rho}^2}~ \left[\mathcal{L}^{(0)} \left(\frac{g_{\rho} \epsilon_i^a \, f^a_i}{m_{\rho}^{3/2}}, \frac{D_{\mu}}{m_{\rho}}, \frac{g_{\rho} H}{m_{\rho}}\right)+\frac{g_\rho^2}{16\pi^2}~\mathcal{L}^{(1)} \left(\frac{g_{\rho} \epsilon_i^a \, f^a_i}{m_{\rho}^{3/2}}, \frac{D_{\mu}}{m_{\rho}}, \frac{g_{\rho} H}{m_{\rho}}\right)+\dots\right]
\end{equation}
where $\lambda_i^a(m_\rho)=g_\rho\epsilon_i^a$, and the $\mathcal{L}^{(n)}$'s are $O(1)$ functions.

The form~(\ref{NDA}) follows from the assumption that the only mass scale of the problem is $m_\rho$ and that all the couplings among the resonances of the flavor sector can be parametrized by a single parameter $g_\rho$. One can equivalently derive~(\ref{NDA}) by first matching the UV theory with a low energy Lagrangian for the composites of masses $\sim m_\rho$. In this case the leading term $\mathcal{L}^{(0)}$ would arise from the tree-level exchange of the resonances, whereas the remainder from loop processes. 

While in generic theories $\mathcal{L}^{(0)}$ already contains all possible operators compatible with the symmetries, it turns out that in all known tractable realizations the resonance spectrum is such that the dipole operators first arise at 1-loop from $\mathcal{L}^{(1)}$. In the following we will assume this is the case.

The spurions $\epsilon^a_i\lesssim1$ measure the amount of compositeness of the field $f_i^a$, and are such that for $\epsilon_i^a\sim1$ the corresponding SM fermion can be interpreted as a fully composite, massless state. 
We will see shortly that the SM mass hierarchy can elegantly arise in theories where the $\epsilon_i^a$'s are hierarchical. One can justify the existence of a hierarchy among the flavor-violating parameters $\epsilon_i^a$ if one postulates that the operators ${\cal O}_i^a$ have large, flavor-dependent scaling dimensions $\Delta_i^a=5/2+\delta_i^a\sim5/2$ at the UV cutoff. 
In this case we expect:
\ba\label{FN}
g_\rho\epsilon_i^a=\lambda^a_i(m_\rho)\sim \lambda^a_i(\Lambda_{\rm UV})\left(\frac{m_\rho}{\Lambda_{\rm UV}}\right)^{\delta_i^a},
\ea
and hence for $\delta_i^a=O(1)$ hierarchical relations can arise in the deep IR even when the $\lambda_i^a$'s are generic, anarchic matrices in the UV. More generally, a controllable explanation of the SM fermion hierarchy can only be given when $\Lambda_{\rm UV}\gg m_\rho$, since when $\Lambda_{\rm UV}\sim m_\rho$ the hierarchy merely represents an assumption on the unknown cutoff theory rather than a prediction of the framework.

In general, also the Higgs doublet should be accompanied by the corresponding ``compositeness" parameter $\epsilon_H$. This quantity does not appear in ${\cal L}_{\rm NDA}$ since we have taken $H$ to be fully composite, and accordingly set $\epsilon_H=1$ in~(\ref{NDA}). 
From a genuinely phenomenological perspective, the assumption of a weakly coupled Higgs at the scale $m_\rho$ would require larger mixing parameters $\epsilon_i^a$ to reproduce the SM masses and would thus lead to larger effects in flavor-violating processes, e.g. in meson-meson mixing. We will emphasize in section~\ref{sec:comp} that in order to avoid dangerous tree-level Higgs corrections to $\Delta F=2$ processes it helps to realize the Higgs as a pseudo Nambu-Goldstone boson (NGB) of the strong sector. 




The NDA Lagrangian~(\ref{NDA}) predicts the following structure for the SM Yukawa matrices of the up and down quarks:
\begin{equation}  \label{eq:yukawas}
(Y_u)_{ij} \sim g_{\rho} \epsilon^q _i \epsilon^u_j,~~~~~~~~~~~~~~~~~(Y_d)_{ij} \sim g_{\rho} \epsilon^q _i \epsilon^d_j.
\end{equation}
(We use $\sim$ throughout the text to indicate that the equalities hold up to unknown $O(1)$ matrices in flavor space.) Eq.~(\ref{eq:yukawas}) suggests that the non-trivial hierarchies of the SM fermion masses could follow from hierarchical mixing parameters $\epsilon_i^a$, as anticipated above. Taking as a phenomenological input $\epsilon^a_1 < \epsilon^a_3 < \epsilon^a_3$, and keeping only the leading terms in the expansion, the Yukawa matrices can be straightforwardly diagonalized by unitary matrices:
\begin{equation}\label{unitary}
(L_u)_{ij} \sim (L_d)_{ij} \sim \min\left(\frac{\epsilon^q_i}{\epsilon^q_j}, \frac{\epsilon^q_j}{\epsilon^q_i}\right)\, , 
\qquad
 (R_{u,d})_{ij} \sim \min\left(\frac{\epsilon^{u,d}_i}{\epsilon^{u,d}_j}, \frac{\epsilon^{u,d}_j}{\epsilon^{u,d}_i}\right).
\end{equation}
The resulting quark masses, renormalized at the scale $m_\rho$, read $m_i^{u,d}=y^{u,d}_iv$, with:
\ba\label{masses}
 (L_u^\dagger Y_u R_u)_{ij}= g_\rho\epsilon^u _i \epsilon^q_i \delta_{ij} \equiv y^u_i \delta_{ij} \, , 
\qquad
 (L_d^\dagger Y_d R_d)_{ij} = g_{\rho} \epsilon^d _i \epsilon^q_i \delta_{ij}  \equiv y^d_i \delta_{ij}\, ,
\ea
and $v(m_Z)\simeq174$ GeV.

Furthermore, noticing that $V_{CKM}=L_d^\dagger L_u\sim L_{u,d}$ we see that the present framework can naturally explain the hierarchical structure of the mixing matrix provided that:
\begin{equation} \label{eq:conditions:eps}
\frac{\epsilon^q_1}{\epsilon^q_2} \sim \lambda~~~~~~~~~~~~~~
\frac{\epsilon^q_2}{\epsilon^q_3} \sim \lambda^2~~~~~~~~~~~~~~
\frac{\epsilon^q_1}{\epsilon^q_3} \sim \lambda^3,
\end{equation}
where $\lambda\simeq0.22$ is the Cabibbo angle. 
In the following we assume that the approximate equalities in~(\ref{eq:conditions:eps}) hold. With these identifications the mixing parameters of the left-handed quarks are completely determined up to an overall normalization factor, whereas the $\epsilon_i^{u,d}$'s are constrained by~(\ref{masses}):
\begin{equation}
\frac{\epsilon_i^{u,d}}{\epsilon_j^{u,d}}=\frac{y_i^{u,d}}{y_j^{u,d}}\frac{\epsilon_{j}^q}{\epsilon_{i}^q}.
\end{equation}
We are thus left with two free parameters that can be $\epsilon_3^q$ and $\epsilon_3^u$ or equivalently one of the two and $g_\rho$. 

The above discussion generalizes to the lepton sector, with the important difference that the neutrinos are much lighter than the charged leptons. As a consequence, it is plausible that the neutrino masses come from a different source, and there is more arbitrariness in the determination of the $\epsilon^a_i$'s.

In fact there is overwhelming experimental evidence indicating that the mixing matrix $V_{PMNS}=L_e^\dagger L_\nu$ is non-hierarchical. Because this latter feature generically occurs whenever $L_\nu$ is anarchic, and whatever the structure of the charged lepton matrix is, we argue that in order to accommodate current data in the lepton sector it suffices to generate hierarchical Yukawa couplings for the charged leptons:
\ba\label{massleptons}
(Y_e)_{ij} \sim g_{\rho} \epsilon^\ell _i \epsilon^e_j,
\ea
and a {\emph{non-hierarchical}} neutrino mass matrix such that $L_\nu=O_{ij}(1)$. 

In models where, in complete analogy with the quark sector, the strong dynamics is responsible for the generation of a Dirac neutrino mass matrix one finds that $L_\nu\sim L_\ell\sim V_{PMNS}$ is the natural prediction. This scenario leads to a complete determination of $\epsilon_i^\ell/\epsilon_j^\ell\sim1$ and $\epsilon_i^e/\epsilon_j^e\sim m^e_i/m^e_j$. 
On the other hand, there is much more freedom in models where the neutrino mass matrix is dominantly generated by couplings involving a SM \emph{bilinear}, rather than the mixing operators of~(\ref{linear}). This program is realized as naturally as in the SM if the neutrinos are Majorana, in which case the leading mass operator would be of the form:
\ba\label{see-saw}
Y_{ij} L_iL_j{\cal O},
\ea
with $Y_{ij}$ anarchic, dimensionless couplings and ${\cal O}$ a composite $SU(2)_L$ triplet operator with scaling dimension $\Delta$. At the scale $\sim m_\rho$, and assuming approximate conformal invariance below the cutoff, Eq.~(\ref{see-saw}) interpolates with:
\ba
\sim Y_{ij}\left(\frac{m_\rho}{\Lambda_{\rm UV}}\right)^{\Delta-2}\frac{L_iH L_jH}{\Lambda_{\rm UV}}
\ea
thus realizing the standard see-saw mechanism when $\Delta\gtrsim2$. Similarly, if the neutrinos are Dirac, by making the mixing parameters $\epsilon^\nu_i$ for the right-handed neutrinos negligibly small, the dominant contribution to the mass matrix would come from higher dimensional operators involving $L_i\nu_j$ at the UV scale~\cite{Agashe:2008fe}. In either case we see that the only constraints imposed on the parameters $\epsilon_i^{\ell,e}$ are given in~(\ref{massleptons}).


Although $\epsilon^\ell_i/\epsilon^\ell_j$ as well as $\epsilon^{e,\ell}_3$ are effectively free-parameters, we will see that the phenomenologically most favorable scenario is that where the left-right and right-left transitions are comparable in magnitude. This is realized when:
\ba\label{R}
\frac{\epsilon_i^\ell}{\epsilon_j^\ell}\sim \frac{\epsilon_i^e}{\epsilon_j^e}\sim\sqrt{\frac{m^e_i}{m^e_j}}.
\ea



\section{Flavor Violation in Composite Higgs Models} 
\label{sec:comp}

We now consider the case of Partial Compositeness in a generic Composite Higgs model. 

An inspection of~(\ref{NDA}) reveals that the main short-distance sources of flavor-violation in these models come from the following $\Delta F=1$ and $\Delta F=2$ operators:
\begin{eqnarray}\label{name}
\mathcal{L}_{\Delta F=1} & \sim & \epsilon_i^{a} \epsilon_j^{b} g_{\rho}
\, \, \frac{v}{m_{\rho}^2} \, \frac{g_{\rho}^2}{(4\pi)^2}
\, \, \overline{f}^{a}_i \sigma_{\mu\nu} g_{\rm SM} F_{\rm SM}^{\mu\nu} f^{b}_j \\\no
 & + & \epsilon_i^{a} \epsilon_j^{b} 
\, \, \frac{g_{\rho}^2}{m_{\rho}^2} 
\, \, \overline{f}^{a}_i \gamma^{\mu} f^{b}_j i H^{\dagger} \overleftrightarrow{D}_{\mu} H \\\no
\mathcal{L}_{\Delta F=2} & \sim & \epsilon_i^{a} \epsilon_j^{b}  \epsilon_k^{c} \epsilon_l^{d} 
\, \,  \frac{g_{\rho}^2}{m_{\rho}^2}
\, \, \overline{f}^{a}_i \gamma^{\mu} f^{b}_j \, \overline{f}^{c}_k \gamma_{\mu} f^{d}_l  
\end{eqnarray}
where $g_{\rm SM} F_{\rm SM}^{\mu\nu}$ is the coupling and field strength of any of the SM gauge groups. (Note that insertions of $g_{\rm SM}F^{\mu\nu}_{\rm SM}=-i[D^\mu,D^\nu]$ are counted like two derivatives and that, as anticipated below~(\ref{NDA}), we assumed that the dipole operators first arise at 1-loop, while the others can be generated at tree level.)

Since our main focus are the dipole operators, we find it convenient to define:
\ba
\Lambda = \frac{4\pi}{g_{\rho}} m_{\rho}.
\ea
Using this definition, and including the $O(1)$ numbers suppressed in~(\ref{name}), we rewrite the above operators as:
\begin{eqnarray}
\mathcal{L}_{\Delta F=1} & = & 
\, \, {\epsilon_i^{a} \epsilon_j^{b} g_{\rho} v}\frac{c^{ab}_{ij, g_{\rm SM}}}{\Lambda^2}~\overline{f}^{a}_i \sigma_{\mu\nu}  g_{\rm SM} F_{\rm SM}^{\mu\nu} f^{b}_j   \label{eq:gen:1}\\
 & + & {\epsilon_i^{a} \epsilon_j^{b} g_{\rho}^2 }\frac{(4\pi)^2}{g_{\rho}^2}\, \frac{c_{ij}^{ab}}{\Lambda^2}  \overline{f}^{a}_i \gamma^{\mu} f^{b}_j~iH^{\dagger} \overleftrightarrow{D}_{\mu} H  \label{eq:gen:1bis}\\
\mathcal{L}_{\Delta F=2} & = &  {\epsilon_i^{a} \epsilon_j^{b}  \epsilon_k^{c} \epsilon_l^{d} g_{\rho}^2}\frac{(4\pi)^2}{g_{\rho}^2}\, \frac{c_{ijkl}^{abcd}}{\Lambda^2} \overline{f}^{a}_i \gamma^{\mu} f^{b}_j \, \overline{f}^{c}_k \gamma_{\mu} f^{d}_l. \label{eq:gen:2}
\end{eqnarray}
Clearly, given $\epsilon_i^{a} \epsilon_j^{b} g_{\rho}$ and $\Lambda$, the case of strong coupling $g_\rho\sim 4\pi$ is phenomenologically the most favorable one.

There are additional contributions to flavor-violation arising from the dimension-6 operators 
\ba
\overline{f}^{a}_i \gamma_{\mu} f^{b}_jD_\nu F_{\rm SM}^{\mu\nu}~~~~~~~{\rm and}~~~~~~~ \overline f^a_i Hf^b_j H^\dagger H. 
\ea
The former lead to FCNC effects that are suppressed by a factor $\sim g_{\rm SM}^2/g_\rho^2$ compared to those induced by those in~(\ref{eq:gen:1bis}), and are hence neglected. On the other hand, the latter generally imply important, long distance contributions to $\Delta F=2$ transitions if the coefficients are not aligned with the Yukawas and the Higgs is light. Indeed, integrating out the Higgs we find that the ratio of the long distance over the short distance contribution~(\ref{eq:gen:2}) scales as:
\ba \label{eq:align}
\frac{g_\rho^4v^4}{m_\rho^2m_h^2}\approx15\left(\frac{g_\rho}{4\pi}\right)^2\left(\frac{10~{\rm TeV}}{\Lambda}\right)^2\left(\frac{125~{\rm GeV}}{m_h}\right)^2.
\ea
Therefore, for $g_\rho \sim 4\pi$ the Higgs exchange generically gives the dominant contribution to meson-meson mixing in these scenarios. 
This problem can be avoided in models where the Higgs is a pseudo-NGB, in which case a careful embedding of the SM fermions in the chiral symmetry of the strong sector forces the Higgs couplings to align with the Yukawa matrix, thus avoiding the above issue~\cite{Agashe:2009di}. In the following we will assume that this mechanism is at play. 
Notice that this alignment will be inevitably spoiled by subleading corrections giving contributions that scale like (\ref{eq:align}) with an extra suppression factor of order $(y_t/4\pi)^2$, and thus under control.

There are also flavor-conserving operators beyond those in~(\ref{name}) that are phenomenologically relevant, such as $H^\dagger W_{\mu\nu}HB^{\mu\nu}$ and $|H^\dagger D_\mu H|^2$. These contribute to the electroweak parameters and will be discussed in Appendix~\ref{pheno}.

\subsection{The CP Asymmetry in D Meson Decays}

Let us now turn to the direct CP asymmetry in D meson decays.
As discussed in \cite{Isidori:2011qw} and \cite{Giudice:2012qq}, the best candidates for producing a sizable effect in the charm sector, while being consistent with the other flavor constraints, are the $\Delta C=1$ chromomagnetic operators. 
Consider then the relevant effective Hamiltonian:
\begin{equation} \label{eq:Q12qu}
{\cal H}_{\rm eff}=\frac{G_F}{\sqrt{2}}\left[C_8\frac{m_c}{4\pi^2}\overline{u_L}\sigma^{\mu\nu}g_sG_{\mu\nu}c_R+C'_8\frac{m_c}{4\pi^2}\overline{u_R}\sigma^{\mu\nu}g_sG_{\mu\nu}c_L\right].
\end{equation}
Matching the above theory with~(\ref{NDA}) at the scale $\mu\sim m_\rho$ gives:
\begin{equation}
C_8(m_\rho)=c_{12,g}^{qu}\frac{4\pi^2\sqrt{2}}{G_F\Lambda^2}\lambda.
\end{equation}
Notice that the coefficient $c_{12,g}^{qu}$ is naturally expected to be of $O(1)$ in our context.
The coefficient $C_8'$ is suppressed by a factor $m_u/m_c\lambda^2\sim5\%$ compared to $C_8$ and will be neglected in our analysis. At leading order in the QCD coupling, running the Wilson coefficient to a scale $\mu<m_\rho$ amounts to taking:
\begin{equation}\label{run}
C_8(\mu)\approx C_8(m_\rho)\times\eta^{\gamma^{(0)}/2\beta_0},
\end{equation}
where $\eta$ is the ratio between the strong coupling at the scale $m_\rho$ and at the scale $\mu$, $\gamma^{(0)}=28/3$ (see, e.g.,~\cite{Degrassi:2005zd}), and $\beta_0=11-2 N_f/3$ is a step function of the number $N_f$ of active flavors. Finally, following~\cite{Isidori:2011qw}\cite{Giudice:2012qq} we write:
\ba\label{GenericACP}
\Delta a_{CP} &\approx& -(0.13\%) \mbox{Im}(\Delta R^{SM}) - 9~ \mbox{Im}(C_8(1~{\rm GeV})) \mbox{Im}(\Delta R^{NP})\\\no
&\approx& -(0.13\%) \mbox{Im}(\Delta R^{SM}) - 0.65\%~\left(\frac{10~ {\rm TeV}}{\Lambda}\right)^2 \left(\frac{\mbox{Im}(c_{12,g}^{qu})}{0.8}\right)~ \frac{\mbox{Im}(\Delta R^{NP})}{0.2},
\ea
where $\Delta R^{SM, NP}$ is the ratio between the hadronic matrix elements of the subdominant SM or new physics operator and the dominant SM contribution. A naive, perturbative estimate gives $\Delta R^{SM}\sim \alpha_s/\pi\sim 0.1$, while perhaps a more conservative one would be $\Delta R^{SM}\sim 1$. The value Im$(\Delta R^{NP})\sim0.2$ was estimated in~\cite{Giudice:2012qq}, where the factorization tools developed in~\cite{Grossman:2006jg} were employed.

\begin{figure}[t]
\begin{center}
\includegraphics[width=0.5\textwidth]{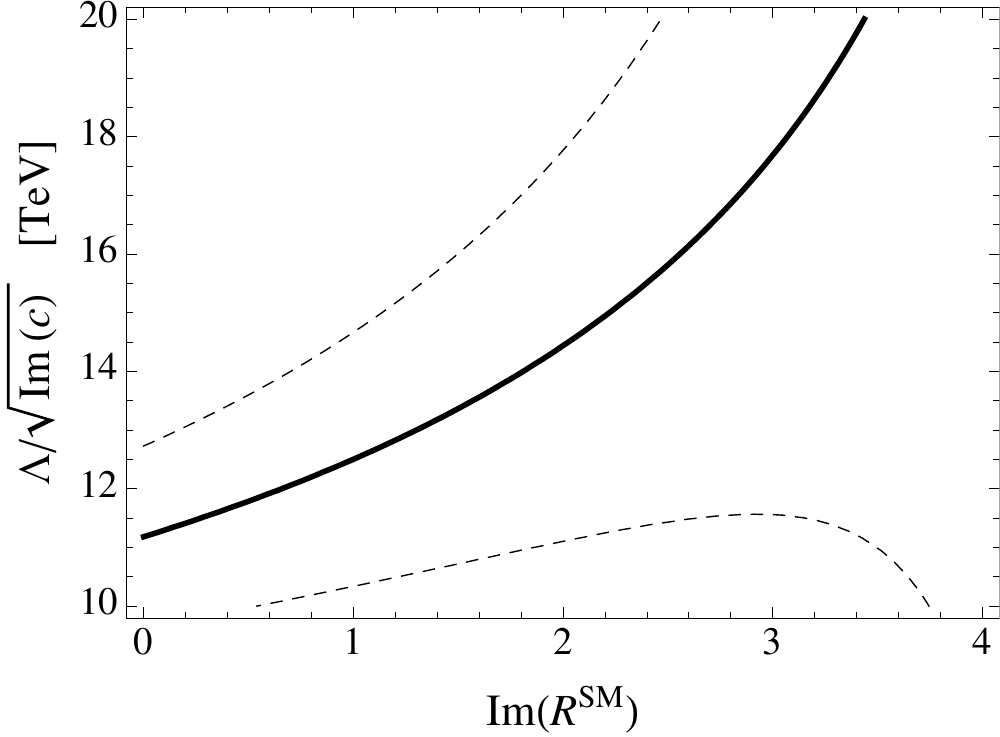}
\caption{\footnotesize{Central value and $1\sigma$ band for the size of the scale $\Lambda/\sqrt{\mbox{Im}(c_{12,g}^{qu})}$ (see definition in Eq.~(\ref{eq:gen:1})) that is needed in order to reproduce (\ref{eq:deltaAcp}) as a function of the SM contribution and taking Im$(\Delta R^{NP})=0.2$, see Eq.~(\ref{GenericACP}).}}
\label{fig:coeffcharm}
\end{center}
\end{figure}


In order to reproduce the observed value (\ref{eq:deltaAcp}) one needs either an enhanced SM contribution $\Delta R^{SM}\sim 5$, or a large new physics contribution, as shown in Fig.~\ref{fig:coeffcharm}. Here we assume that $\Delta a_{CP}$ is saturated by physics beyond the SM, and choose:
\begin{equation} \label{eq:fixingthescale}
 \Lambda = 10 \mbox{ TeV}, 
\quad  \quad~~~~~~~~~
\mbox{Im}(c^{qu}_{12,g}) \sim 1 \, .
\end{equation}
It is important to emphasize that our conclusions will necessarily be affected by the large systematic uncertainty associated with the long distance QCD effects encoded in the matrix elements $\Delta R^{SM,NP}$.

Having fixed the scale $\Lambda$ according to (\ref{eq:fixingthescale}), we then consider the bounds on the dimensionless coefficients $c_{ij, {\rm SM}}^{ab}$, $c_{ij}^{ab}$, and $c_{ijkl}^{abcd}$ (evaluated at the scale  $\sim m_\rho$) of the other flavor-violating operators~(\ref{eq:gen:1})--(\ref{eq:gen:2}). 
We summarize our results in Table \ref{table:generic}, referring to Appendix \ref{pheno} for details. The reader should keep in mind that the coefficients of~(\ref{eq:gen:1})--(\ref{eq:gen:2}) have NDA values $O(1)$, therefore values much smaller than one can only be accounted for by additional, non-generic assumptions on the UV dynamics.

For the case of quark flavor violation, the compatibility with the experimental data is basically guaranteed by the structure~(\ref{eq:gen:1})--(\ref{eq:gen:2}) for a maximally strong sector $g_\rho\sim4\pi$. Some moderate suppressions however are needed in the observables $\epsilon_K$ and $\epsilon'/\epsilon$, while a stronger one is necessary in the flavor-blind CP-violating coefficients of the neutron Electric Dipole Moments (EDM). Perhaps the most significant constraint is the one on the operator $\overline{u}\sigma^{\mu\nu}g_sG_{\mu\nu}u_{L,R}$ as its structure is identical -- except for being flavor conserving -- to the one used to produce the effect in $\Delta a_{CP}$.
In fact the strong bound on the analogous operator made of down-type quarks suggests that in a concrete model one needs to suppress all the operators involving down-type quarks, thus relaxing also the bounds from $\epsilon_K$ and $\epsilon'/\epsilon$.

In conclusion, a robust prediction of this framework is that the neutron EDM, together possibly with $\epsilon_K$, $\epsilon'/\epsilon$ as well as $B\to X_s\gamma$, should be close to the current experimental sensitivity.
Notice also that new physics effects are predicted in the process $K^+\to\pi^+\bar\nu\nu$ to be within the  planned sensitivity of the NA62 experiment \cite{NA62}, see (\ref{eq:KtoPiNuNu}) in Appendix \ref{pheno}.

As we explain in Appendix \ref{pheno} the LHC will not be able to detect any sizable deviation in neutral flavor changing decays of the top quark. With $m_\rho=10$~TeV and order one coefficients the leading effect, to be expected in the $t\to cZ$ branching fraction, is roughly one order of magnitude below the foreseen experimental sensitivity.

The minimal scenario under consideration is {\emph{not}} compatible with the stringent experimental data in the lepton sector\footnote{See also \cite{Agashe:2006iy}, that corresponds to $g_\rho \sim 1$ in our case.}. While no tension in $\mu \rightarrow e$ transitions in nuclei nor in $(g-2)_\mu$ is observed\footnote{We cannot explain the observed 3.1$\sigma$ $(g-2)_\mu$ anomaly \cite{gimeno2} due to the large $\Lambda$ value needed to accommodate the LHCb result.}, the unnaturally small coefficients required to satisfy the electron EDM and the $\mu\to e\gamma$ bounds strongly point towards less minimal scenarios.

If we momentarily set aside the problems associated with the lepton sector 
we conclude that models of Partial Compositeness with $m_\rho\sim10$ TeV represent phenomenologically viable and theoretically compelling theories of (quark) flavor. Furthermore, as discussed below~(\ref{eq:gen:2}), scenarios of Partial Compositeness are especially well motivated if the Higgs doublet arises as a pseudo-NGB of the new strong flavorful dynamics, thus suggesting a non-trivial link between the flavor puzzle and the weak scale. 
In this respect notice that, in the absence of new dynamical assumptions, the Higgs potential in composite Higgs models is dominantly determined by the top-quark couplings and approximately reads:
\ba\label{potential}
V(H)\sim\frac{3}{4\pi^2}(\epsilon_3^{q,u})^2m_\rho^4\;\overline V\left(\frac{g_\rho H}{m_\rho}\right).
\ea
The natural vacuum therefore sits at $v\sim m_\rho/g_\rho$, so that  to obtain a phenomenologically viable model one needs at least a fine-tuning of order:
\ba
\frac{g_\rho^2v^2}{m_\rho^2}\approx5\%\left(\frac{10~{\rm TeV}}{\Lambda}\right)^2 \, .
\ea
In general, however, a stronger tuning is required to obtain a light physical Higgs. After the electroweak vacuum has been set to its phenomenological value, from~(\ref{potential}) 
we find that a tuning between the percent and the permille level is needed to accommodate $m_h\sim125$ GeV.
If this is really how nature works, then ATLAS and CMS will not be able to directly probe the confinement scale $\sim m_\rho$, and the most striking, generic signatures of Partial Compositeness would be visible only in indirect, precision measurements.

This conclusion would change if the typical mass $m_\psi$ of the fermionic resonances of the new sector is somewhat smaller than $m_\rho$, in which case the fine-tuning problem can be ameliorated, as recently discussed in \cite{Matsedonskyi:2012ym}-\cite{Marzocca:2012zn}. 

Let us briefly see how power counting should be modified under this assumption. Formally, the fermion resonances $\psi$ can be made parametrically lighter than the dynamical scale by imposing an approximate chiral symmetry. As a consequence, all operators in~(\ref{NDA}) that violate such a symmetry should be accompanied by appropriate powers of the small parameter 
\ba
\frac{g_\psi}{g_\rho}\equiv \frac{m_\psi}{m_\rho}\ll1. 
\ea
Terms that violate the chiral symmetry include the mass mixing with the SM fermions, now controlled by the operators $\tilde\epsilon_i^a m_\psi \overline{\psi}^a_if_i^a P(g_\rho H/m_\rho)$, and the coupling of the $\psi$'s to the Higgs doublet, which are now proportional to $g_\psi$ rather than to $g_\rho$. As a result one finds that the Yukawa matrix scales as $\sim g_\psi \tilde\epsilon_i^a\tilde\epsilon_j^b$, and similarly that the Higgs boson mass is reduced by some power of $g_\psi/g_\rho$ compared to the generic case. 

However, no suppression is expected for non-chiral couplings among the $\psi$'s and the other heavy resonances, which are still set by $g_\rho$. This implies that chirally-invariant flavor-violating operators will become parametrically more relevant than in the generic $g_\psi\sim g_\rho$ case. Keeping the Yukawa matrix as well as $\Lambda$ fixed, we consistently find that the couplings of the operators in~(\ref{eq:gen:1}) are unchanged, but those in~(\ref{eq:gen:1bis}) and~(\ref{eq:gen:2}) are parametrically \emph{enhanced} by a factor $g_\rho/g_\psi$ and $(g_\rho/g_\psi)^2$, respectively. We find this 
unsatisfactory for our purpose, and therefore do not discuss this regime any further.

\begin{table}[t]
\begin{center}
\begin{tabular}{c|c c|c} \hline\hline
\rule{0pt}{1.2em}%
Operator $\Delta F=2$ &
 Re$(c) \times (4\pi / g_{\rho})^2 $ & Im$(c) \times (4\pi / g_{\rho})^2 $ 
& Observables \\
 \hline \hline
$(\bar s_L \gamma^\mu d_L )^2$   &
$6 \times 10^2 \,\,  \left(\frac{\epsilon_3^u }{\epsilon_3^q}\right)^2 $& $ 2 \,\,\left(\frac{\epsilon_3^u }{\epsilon_3^q  }\right)^2$ &
 $\Delta m_K$; $\epsilon_K$~\cite{Bona:2007vi}\cite{Isidori:2010kg}  \\
 $(\bar s_R d_L)^2$   &
$ 500  $& $ 2 $ &
 " \\
($\bar s_R\, d_L)(\bar s_L d_R$)   &
 $2 \times 10^2 \,\,$& $ 0.6$ &
 " \\
\hline $(\bar c_L \gamma^\mu u_L )^2$  &
$4 \times 10^2 \,\, \left(\frac{\epsilon_3^u }{\epsilon_3^q }\right)^2$& $ 70 \,\, \left(\frac{\epsilon_3^u }{\epsilon_3^q }\right)^2$ &
 $\Delta m_D$; $|q/p|, \phi_D$~\cite{Bona:2007vi}\cite{Isidori:2010kg} \\
$(\bar c_L\, u_R)^2$   &
 $30 $& $ 6 \,\, $ &
"\\
($\bar c_R\, u_L)(\bar c_L u_R$)   &
 $3 \times 10^2 $& $ 50 \,\, $ &
"\\
\hline$(\bar b_L \gamma^\mu d_L )^2$    &
  $5 \,\, \left(\frac{\epsilon_3^u }{\epsilon_3^q }\right)^2$ & $ 2 \,\, \left(\frac{\epsilon_3^u }{\epsilon_3^q  }\right)^2$ &
 $\Delta m_{B_d}$; $S_{\psi K_S}$~\cite{Bona:2007vi}\cite{Isidori:2010kg}  \\
($\bar b_R\, d_L)^2$  &  
 $80  \,\,$ & $ 30 \,\, $ &  
" \\
($\bar b_R\, d_L)(\bar b_L d_R)$  &  
 $3 \times 10^2  \,\,$ & $ 80 \,\, $ &  
" \\
\hline $(\bar b_L \gamma^\mu s_L )^2$    &  \multicolumn{2}{c|}{$ 6 \,\, \left(\frac{\epsilon_3^u }{\epsilon_3^q }\right)^2$} &
  $\Delta m_{B_s}$~\cite{Bona:2007vi}\cite{Isidori:2010kg} \\
($\bar b_R \,s_L)^2$  &   \multicolumn{2}{c|}{$1 \times 10^2  \,\,$}   &
" \\
($\bar b_R \,s_L)(\bar b_L s_R)$  &   \multicolumn{2}{c|}{$3 \times 10^2  \,\,$}   &
" \\  
  \hline\hline
  
 \rule{0pt}{1.2em}%
Operator $\Delta F=1$ & 
 Re$(c) $ & Im$(c) $ 
& Observables \\
    \hline\hline
    $\overline{s_R}\sigma^{\mu\nu}eF_{\mu\nu}b_L$    &  \multicolumn{2}{c|}{$ 1 $} &
  $B\to X_s$~\cite{Altmannshofer:2011gn} \\
  $\overline{s_L}\sigma^{\mu\nu}eF_{\mu\nu}b_R$    &  $ 2 $ & $9$ &
  " \\
  \hline
  $\overline{s_R}\sigma^{\mu\nu}g_sG_{\mu\nu}d_L$    &  - &$ 0.4 $ &
  $K\to 2\pi$; $\epsilon'/\epsilon$~\cite{Gedalia:2009ws} \\
  $\overline{s_L}\sigma^{\mu\nu}g_sG_{\mu\nu}d_R$    &  - &$ 0.4 $&
 " \\
 \hline
  $\bar s_L\gamma^\mu b_L\, H^\dagger i\overleftrightarrow D_\mu H$    &   \multicolumn{2}{c|}{$30\left(\frac{g_\rho}{4\pi}\right)^2(\epsilon_3^u)^2$}  &
 $B_s\to \mu^+\mu^-$ \cite{LHCbLaThuile} \\  
  $\bar s_L\gamma^\mu b_L\, H^\dagger i\overleftrightarrow D_\mu H$    &   $6~ \left(\frac{g_\rho}{4\pi}\right)^2(\epsilon_3^u)^2$ & $10~ {\left(\frac{g_\rho}{4\pi}\right)^2}(\epsilon_3^u)^2$  &
 $B\to X_s \, \ell^+\ell^-$ \cite{Altmannshofer:2011gn}  \\  
  $\bar s_L\gamma^\mu d_L\, H^\dagger i\overleftrightarrow D_\mu H$    &  -  & $3~ {\left(\frac{g_\rho}{4\pi}\right)^2}(\epsilon_3^u)^2$  &
 $\epsilon '/\epsilon$ \cite{Bauer:2009cf}  \\ 
   \hline\hline

 \rule{0pt}{1.2em}%
Operator $\Delta F=0$ & 
 Re$(c) $ & Im$(c) $ 
& Observables \\
    \hline\hline  
  $\overline{d}\sigma^{\mu\nu}eF_{\mu\nu}d_{L,R}$    &  - & {$ 3\times10^{-2} $} &
  neutron EDM~\cite{Baker:2006ts}\cite{Pospelov:2000bw} \\
  $\overline{u}\sigma^{\mu\nu}eF_{\mu\nu}u_{L,R}$    &  - & $ 0.3 $ &
  " \\  
  $\overline{d}\sigma^{\mu\nu}g_sG_{\mu\nu}d_{L,R}$    &  - & $ 4\times10^{-2} $ &
  " \\
  $\overline{u}\sigma^{\mu\nu}g_sG_{\mu\nu}u_{L,R}$    & - & $ 0.2 $ &
  " \\
 \hline
  $\bar b_L\gamma^\mu b_L\, H^\dagger i\overleftrightarrow D_\mu H$    &   \multicolumn{2}{c|}{$5\left(\frac{g_\rho}{4\pi}\right)^2(\epsilon_3^u)^2$}  &
 $Z \rightarrow b \bar b$ \cite{ALEPH:2005ab} \\ 

  \hline\hline
 \rule{0pt}{1.2em}%
Leptonic Operator & 
 Re$(c) $ & Im$(c) $ 
& Observables \\
    \hline\hline  
  $\overline{e}\sigma^{\mu\nu}eF_{\mu\nu}e_{L,R}$    &  - & {$ 8\times10^{-3} $} &
  electron EDM~\cite{Regan:2002ta} \\
  \hline
   $\overline{\mu}\sigma^{\mu\nu}eF_{\mu\nu}e_{L,R}$    &  \multicolumn{2}{c|}{$ 4\times10^{-3}$} &
  $\mu\to e\gamma$~\cite{Adam:2011ch}  \\
   \hline
  $\bar e\gamma^\mu \mu_{L,R}\, H^\dagger i\overleftrightarrow D_\mu H$    &   \multicolumn{2}{c|}{$ 1.5\left(\frac{g_\rho}{4\pi}\right)\frac{\epsilon_3^e}{\epsilon_3^\ell}$}  &
 $\mu(Au)\to e(Au)$ \cite{lfv} \\  

   \hline\hline
\end{tabular}

\caption{\label{table:generic}\footnotesize{Upper bounds on the dimensionless coefficients of the operators in the notation (\ref{eq:gen:1})--(\ref{eq:gen:2}), with $\Lambda=4\pi m_\rho/g_\rho=10$ TeV. The bound is on the coefficients renormalized at 10 TeV, and we report the strongest ones. To minimize the constraints in the lepton sector we assumed~(\ref{R}).
Notice that the combinations  $g_\rho \epsilon_3^q \epsilon_3^u$ and $g_\rho \epsilon_3^\ell \epsilon_3^e$ are fixed to be respectively $y_t$ and $y_\tau$.
The experimental bounds are taken from the references in the third column. 
See Appendix~\ref{pheno} for details.
}
}
\end{center}
\end{table}

\clearpage


\section{Partial Compositeness and Supersymmetry}
\label{sec:susy}

To reconcile Partial Compositeness with the stringent bounds in the lepton sector one could either introduce additional assumptions on the UV theory, or take a more phenomenological approach and explain the lepton hierarchy within a different framework. These alternatives however are not completely satisfactory from a theoretical point of view, as in doing so one abandons the simplicity of the minimal construction. 
In this Section we instead propose to combine the paradigm of Partial Compositeness with Supersymmetry (SUSY), as first done in~\cite{flavorful1} (for related work see also~\cite{Davidson:2007si}\cite{Dudas:2010yh}). Intriguingly, we find that these \emph{flavorful SUSY} models can generate the observed effect in $\Delta a_{CP}$ while being roughly consistent with the other data.

The basic assumptions are the same as in Section~\ref{framework}, but now replacing the SM fields $f^a_i$ with the corresponding chiral superfields $\Phi^a_i$. To clarify the general picture, let us call $\Lambda_{S}$ the scale at which SUSY-breaking is communicated to the visible sector, while $m_\rho\equiv\Lambda_F$ denotes now the scale below which the flavor structure is completely encoded in the Yukawa matrices, given in Eqs.~(\ref{eq:yukawas}) and~(\ref{massleptons}). 

At the scale $\Lambda_S$ we invoke a mechanism like gauge mediation or some friendly string construction to generate flavor universal soft terms for the light fields. On the other hand we don't need to commit to any specific structure for the soft terms in the heavy sector.%

We then require $\Lambda_S\gg m_\rho$. This has to be contrasted with the usual case in gauge mediation where $m_{\rho}\equiv \Lambda_F \gg \Lambda_{S}$, in which universality of the soft terms at low energy is ensured  by communicating supersymmetry breaking to the visible sector at a scale where the flavor dynamics is already locked in the Yukawa matrices. 

Above $\Lambda_F$ the interactions among the heavy fields are flavor anarchic. Several  options can be imagined for the flavor sector, but  it is reasonable to expect that its soft terms will be $O(1)$ non-universal at the scale $m_\rho$. A first case is when $g_\rho\sim 4\pi$ at $m_\rho$: there quantum effects are clearly distorting any original 
universality. In particular that  ensures that the resulting effective $A$-terms are sizable and not exactly aligned with the low-energy Yukawa matrices. In a  second case  $g_\rho<4\pi$, but RG effects still generate $O(1)$ non-universality thanks to a sufficiently large  separation between $\Lambda_S$ and $m_\rho$ :  $ (g_\rho/4\pi)^2\log\Lambda_S/ m_\rho=O(1)$. Finally a third conceivable situation is one where the soft terms in the flavor sector are already generic at the scale $\Lambda_S$. In all the above cases, by integrating out the flavor sector at the scale $m_\rho$, there arise corrections to soft terms that are  generic and  controlled by the $\epsilon$'s.

In conclusion the MSSM soft terms are expected to be of the general form:
\ba\label{soft}
(m^2_{a})_{ij}&\sim&\tilde m^2_a\delta_{ij}+\epsilon^a_i\epsilon^a_j c^a_{ij}\tilde m_0^2,\quad a=q,u,d,\\\no
A^{u,d}_{ij}&\sim&g_\rho \epsilon^q_i\epsilon^{u,d}_j d^{u,d}_{ij} A_0,
\ea
where $c$ and $d$ are order one coefficients. Similar expressions hold for the sleptons. Strictly speaking we have this structure at the scale $m_{\rho}$, and there are $O(1)$ modifications from the RG evolution down to low energy. These effects however are small for the first two generations, which play the central role in our analysis, so for our purposes it is enough to assume (\ref{soft}) to hold at low energies.

To be more specific, we can parametrize SUSY-breaking by the vev of a single spurion chiral superfield $X=\theta^2 \tilde{m}_0$ and, working at leading order in a derivative expansion, consider the following NDA Lagrangian:
\ba\label{NDAsusy}
{\cal L}^{\rm SUSY}_{\rm NDA}&=&\int d^2\theta \int d^2\overline{\theta}~ \frac{m_\rho^2}{g_\rho^2}~{\cal K}\left(\frac{\epsilon^a_ig_\rho\Phi^a_i}{m_\rho},X,\frac{g_\rho H_{u,d}}{m_\rho}\right)\\\no
&+&\left[\int d^2\theta ~\frac{m_\rho^3}{g_\rho^2}~{\cal W}\left(\frac{\epsilon^a_ig_\rho\Phi^a_i}{m_\rho},X,\frac{g_\rho H_{u,d}}{m_\rho}\right)+{\rm h.c.}\right].
\ea
Here ${\cal K}$ (${\cal W}$) is a generic $O(1)$ real (holomorphic) function. Note that we assumed that the Higgses $H_{u,d}$ are fully composite\footnote{It is straightforward to generalize our results to the case of partially composite Higgses.}, in analogy with the generic models of section~\ref{framework}.
From the above Lagrangian we can easily derive the soft masses of the squarks and sleptons and the A-terms, which are the main source of flavor violation within the MSSM. An inspection of~({\ref{NDAsusy}) shows that the former $\propto\tilde m_0^2$ arise entirely from the Kahler, whereas the latter $\propto A_0$ receive contributions also from the superpotential. Because of this we will treat $\tilde m_0^2$ and $A_0\propto\tilde m_0$ as distinct parameters.
Schematically, the flavorful SUSY breaking terms read $m_{ij}^2\sim\epsilon_i\epsilon_j\left(1+\epsilon_k\epsilon_k\right)\tilde m_0^2\sim\epsilon_i\epsilon_j\tilde m_0^2$ and $A_{ij}\sim g_\rho\epsilon_i\epsilon_j\left(1+\epsilon_k\epsilon_k\right)A_0\sim g_\rho\epsilon_i\epsilon_jA_0$. Finally, including possible ``direct" positive contributions, we find that the relevant soft parameters have the form (\ref{soft}).

Our primary goal is to investigate the viability of the minimal scenarios in which the \emph{only} source of flavor violation is parametrized by the spurions $\epsilon^a_i$'s. For this reason, as well as simplicity, we will focus on models with non-hierarchical soft SUSY masses for squarks and sleptons:
\ba\label{degenerate}
\tilde m^2_{q,u,d}\sim\tilde m^2_{\ell,e} \sim\tilde m^2 \, . 
\ea
More general cases can certainly be considered, but we believe that a qualitative understanding of the actual viability of SUSY scenarios of Partial Compositeness can already be obtained assuming an approximately degenerate spectrum. One could for example introduce hierarchies among the squark and slepton masses, as it happens in minimal gauge mediation. The interested reader can translate our results to these cases with minimal effort. One could also consider scenarios where additional sources of flavor violation are introduced. An interesting case that attracted renewed attention after the latest LHC direct limits on the superpartners is the one where the third quark generation is lighter than the first two (see~\cite{Brust:2011tb} for an effective description). 
In the latter case flavor-violation will be mainly controlled by the alignment in the first two families rather than by the $\epsilon_i^a$'s. 



Before analyzing in detail the bounds on the present model, it is instructive to compare the flavor-violating structure following from~(\ref{soft}) to that of the non-SUSY model~(\ref{eq:gen:1})--(\ref{eq:gen:2}) studied in the previous section. 
An important difference between the two scenarios is that in the present case we are free to decouple the strong flavor sector by taking:
\ba
m_\rho\gg10~{\rm TeV} 
\ea
without introducing additional naturalness issues. Under this working hypothesis, flavor-violating contributions will mainly arise through loops of the SM superpartners, and will hence be controlled by the SM couplings and the superpartner mass scale $\tilde m$. We thus define $\Lambda = 4\pi \tilde{m}/ g_{\rm SM}$, where $g_{\rm SM}$ collectively denotes a SM coupling, and find that the dominant source of flavor-violation follows from:
\ba\label{eq}
\mathcal{L}_{\Delta F=1} & \sim & \, \, {\epsilon_i^{a} \epsilon_j^{b} g_{\rho} v}\frac{1}{\Lambda^2}\, \, \overline{f}^{a}_i \sigma_{\mu\nu}  g_{\rm SM} F_{\rm SM}^{\mu\nu} f^{b}_j  \,\\\no
 & + & 
\, \, {\epsilon_i^{a} \epsilon_j^{b} g_{\rho}^2 } \frac{g_{\rm SM}^2}{g_{\rho}^2}\, \frac{1}{\Lambda^2} \, \,\overline{f}^{a}_i \gamma^{\mu} f^{b}_j i H^{\dagger} \overleftrightarrow{D}_{\mu} H  \,\\\no
\mathcal{L}_{\Delta F=2} & \sim &  
\, \,  {\epsilon_i^{a} \epsilon_j^{b}  \epsilon_k^{c} \epsilon_l^{d} g_{\rho}^2} \frac{g_{\rm SM}^2}{g_{\rho}^2}\, \frac{1}{\Lambda^2} 
\, \, \overline{f}^{a}_i \gamma^{\mu} f^{b}_j \, \overline{f}^{c}_k \gamma_{\mu} f^{d}_l 
\ea
For a fixed $\Lambda$, the above structure is formally identical to~(\ref{eq:gen:1})--(\ref{eq:gen:2}) provided one replaces $4\pi/g_{\rho}\to g_{\rm SM}/g_{\rho}$. The two realizations will therefore lead to comparable predictions if we set $g_\rho\sim4\pi$ in the models of Section~\ref{framework} and assume $g_\rho \sim g_{\rm SM}$ in the SUSY model. More generally, however, we expect that when $g_{\rho} \gg  g_{\rm SM}$ in~(\ref{eq}) the flavor violating processes mediated by $\Delta F=2$ and penguin operators will be parametrically suppressed compared to the non-SUSY case. 

To quantitatively assess flavor violation in the present model we find it convenient to employ the mass insertion approximation in the notation of \cite{Gabbiani:1996hi}. We thus rotate the superfields into the basis where the Yukawas are diagonal via the unitary matrices~(\ref{unitary}). The structure one obtains is still the same as in~(\ref{soft}), and we define:
\begin{eqnarray}\label{massinse}
(\delta_{ij}^{u,d})_{LL} = (c^{u,d}_{ij})_{LL} \,\, \times \,\, \frac{\tilde{m}_0^2}{\tilde{m}^2} \,\,  \epsilon^{q}_i \epsilon^{q}_j,   &&
(\delta_{ij}^{u,d})_{RR} = (c^{u,d}_{ij})_{RR} \,\,\times\,\,\frac{\tilde{m}_0^2}{\tilde{m}^2} \,\, \epsilon^{u,d}_i \epsilon^{u,d}_j,    \nonumber \\
(\delta_{ij}^{u,d})_{LR} = (c^{u,d}_{ij})_{LR}  \,\,\times\,\,  g_{\rho}  \,\epsilon^{q}_i \epsilon^{u,d}_j  \, \frac{v_{u,d} \, A_0}{\tilde{m}^2},     &&
(\delta_{ij}^{u,d})_{RL} = (c^{u,d}_{ij})_{RL}  \,\,\times\,\,  g_{\rho}  \,\epsilon^{u,d}_i \epsilon^{q}_j   \, \frac{v_{u,d} \, A_0}{\tilde{m}^2},   \label{eq:deltas:susy}
\end{eqnarray}
where $v_u = v \sin\beta$ and $v_d=v \cos\beta$ with $v\approx174$ GeV, while $(c^{f}_{ij})_{LR,\, RL}$ are coefficients with NDA value $O(1)$. Analogous expressions hold for the lepton sector, and will not be reproduced here for brevity. 

In writing~(\ref{eq:deltas:susy}) we neglected the contribution of the $\mu$ term in the family-diagonal $LR$ insertions, which can be relevant when $\tan\beta$ is large even for $\mu<A_0$. Although these terms can be phenomenologically relevant, for example in setting bounds on the phase of the $\mu$ term times gaugino masses, we do not consider them here since they do not provide direct constraints on the coefficients $(c^f_{ij})_{LR,RL}$ defined in (\ref{eq:deltas:susy}). 

In the regime $g_\rho\gg1$ 
the largest chirality-violating contributions always come from single $(\delta_{ij})_{LR}$ insertions, while the triple insertions of the type $(\delta_{i3})_{LL} (\delta_{33})_{LR} (\delta_{3j})_{RR}$ provide subleading corrections. In the opposite limit, $g_\rho\sim y_t$, one requires $\epsilon_3^{q,u}\sim 1$ in order to reproduce the correct top mass value, and finds that the diagonal top squark masses receive an unsuppressed flavor-violating contribution,
\begin{equation}\label{splitfamilies}
\left(m^2_{q,u}\right)_{33}\sim \tilde m^2+ \tilde m_0^2,
\end{equation}
which can either add up or cancel against the flavor universal one. This possibility can be used to realize the so called \emph{split-families} ansatz in a corner of the parameter space of flavorful SUSY. In this limit $(\delta_{33}^u)_{LL(RR)}$ are $ O (1)$ while $(\delta_{33}^u)_{LR}$ is still smaller than $1$ as long as $\left(m_{q,u}\right)_{33}>m_t$. In order to make the mass insertion approximation well-defined it is thus sufficient to define the stop propagators including the full $m_{33}^2$ mass in eq.\eqref{splitfamilies} and not just the flavor universal term. In this way the former conclusions about the hierarchy between single and multiple insertions are still valid, and we only expect $O(1)$ modifications in our numerical results.


As argued below Eq.~(\ref{eq}), and recently emphasized in~\cite{Giudice:2012qq}, left-right transitions will play an important role in our analysis. Interestingly, Partial Compositeness implies that their off-diagonal structure is automatically the same as that of the Yukawa's, though not necessarily aligned, such that we can write $A^{u,d}_{ij}\propto m^{u,d}_j\epsilon_i^q/\epsilon_j^q$. In this sense, the present framework can be seen as a concrete realization of the scenario dubbed `disoriented A-terms' in~\cite{Giudice:2012qq}, and explicitly designed to generate the observad $\Delta a_{CP}$ in the D meson decays. 

\subsection{The CP Asymmetry in D Meson Decays}

\begin{figure}[t]
\begin{center}
\includegraphics[width=0.6\textwidth]{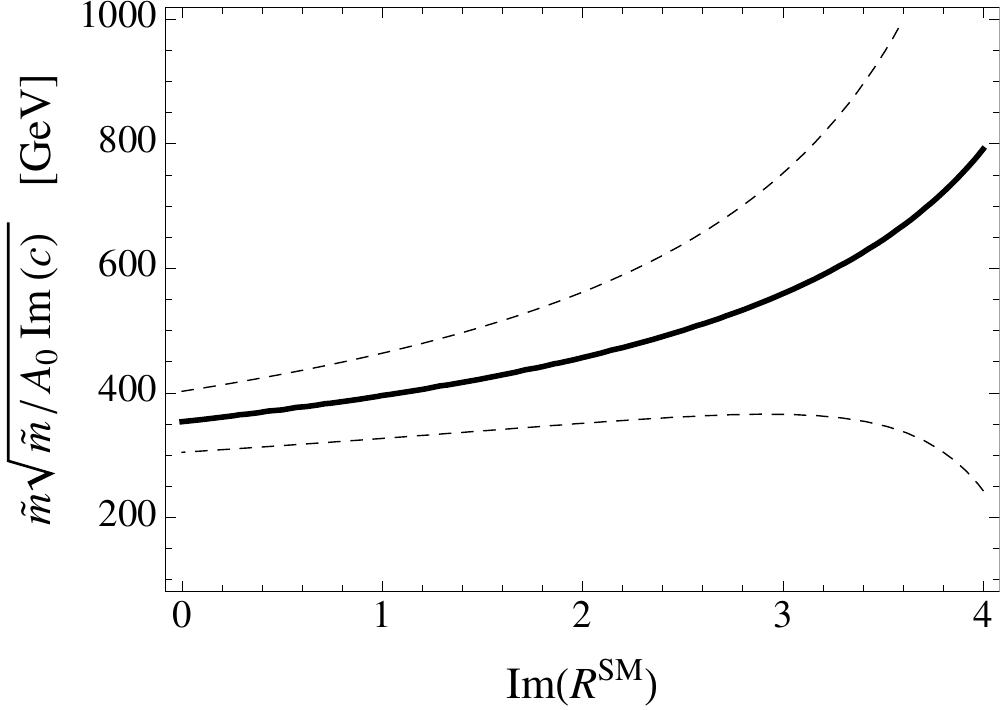}
\caption{\footnotesize{Central value and $1\sigma$ band for the size of $\tilde m\sqrt{\tilde m/A_0 {\rm Im}(c_{12}^{u})_{LR}}$ required in order to reproduce (\ref{eq:deltaAcp}) as a function of the SM contribution and taking Im$(\Delta R^{NP})=0.2$, see Eq.~(\ref{SUSYACP}).}}
\label{fig:coeffcharmSUSY}
\end{center}
\end{figure}

We are now ready to see which region of the parameter space is favored by the LHCb result. 
The dominant contribution to the direct CP asymmetry~(\ref{eq:deltaAcp}) is again induced by the chromomagnetic operator of (\ref{eq:Q12qu}). Analogously to the non-SUSY scenario, the coefficient $C_8'$ is suppressed by a factor $(\delta_{12}^u)_{RL}/(\delta_{12}^u)_{LR}\sim m_u/m_c\lambda^2\sim5\%$ compared to $C_8$ and can hence be neglected. The relevant contribution to the Wilson coefficient at the scale $\tilde m$ reads:
\ba
C_8(\tilde m)&=&\frac{\sqrt{2}}{G_F}\frac{\alpha_s\pi}{2m_c}\frac{m_{\tilde g}}{m_{\tilde q}^2}\left(-\frac{1}{3}M_1(x)-3M_2(x)\right)(\delta_{12}^u)_{LR}\\\no
&\approx&\frac{\sqrt{2}}{G_F}\frac{\alpha_s\pi}{\tilde m^2}\left(-\frac{5}{36}\right)(c_{12}^{u})_{LR}\frac{\lambda A_0}{\tilde m}
\ea
where the loop functions $M_{1,2}(x)$ are defined in \cite{Gabbiani:1996hi}.
In the above expression $x=m_{\tilde g}^2/m_{\tilde q}^2$, whereas in the approximate equality we took $x=1$ and plugged~(\ref{massinse}) in. Including the QCD running we finally obtain:
\ba\label{SUSYACP}
\Delta a_{CP} \approx -(0.13\%) \mbox{Im}(\Delta R^{SM}) - 0.65\%~\left(\frac{1~{\rm TeV}}{\tilde m}\right)^2 \left(\frac{A_0}{8\tilde m}\;\mbox{Im}(c_{12}^{u})_{LR}\right)~ \frac{\mbox{Im}(\Delta R^{NP})}{0.2}.
\ea
Following the same logic of Section~\ref{sec:comp} we assume that $\Delta a_{CP}$ is saturated by physics beyond the SM. From~(\ref{SUSYACP}) we see that this requires:
\begin{equation}  \label{eq:susy:assumptions}
\mbox{Im}(c_{12}^{u})_{LR}\times\frac{A_0}{\tilde{m}}\times\left(\frac{1~{\rm TeV}}{\tilde m}\right)^2\sim8,
\end{equation}
modulo hadronic uncertainties involved in estimating $\Delta R^{SM,NP}$. A more favorable condition is met if we consider a nondegenerate spectrum for the sfermions. However, the general trend is not expected to change dramatically, so we stick to the minimal choice~(\ref{degenerate}).

Taking $\tilde m\sim \tilde m_0 \sim1$ TeV allows us to marginally evade the current direct bounds on the sparticle spectrum. However, for ${\rm Im}(c_{12}^{u})_{LR}\sim1$, this possibility requires $A_0/\tilde m>3$ and generally implies the existence of new vacua with color and electromagnetic quantum numbers~\cite{CCB}. This pathology is avoided if the coefficient $(c_{12}^{u})_{LR}$ turns out to be larger than its NDA estimate, such that even for $A_0\sim\tilde m$ the condition~(\ref{eq:susy:assumptions}) can be satisfied\footnote{In our case the vacuum stability bound on the non-diagonal combination $(c_{12}^{u})_{LR}\times\frac{A_0}{\tilde{m}}$ is roughly 15, thanks to the $\lambda$ suppression of the $A$-term.}. Alternatively, keeping all coefficients at their natural values we could impose $A_0/\tilde m<3$ and find that $\tilde{m}\lesssim600$ GeV is now necessary. This choice can be consistent with the LHC bounds only if the typical signature of our framework departs significantly from the vanilla MSSM scenario with R-parity. Interestingly, the possibility of having sizable R-parity violating couplings turns out to be viable, as shown in section~\ref{RPVsection} below. 
As a side note $\tilde m \sim 1$ TeV and $A_0/\tilde m\sim 2-3$ are in the right ballpark to obtain a Higgs boson mass of 125 GeV without going beyond the MSSM at low energies, see e.g.~\cite{Hall:2011aa}.

Referring to Appendix \ref{SUSYconstraints} for details, we repeat the analysis of the flavor bounds of Section \ref{sec:general}. Our results are summarized in Table \ref{table:susy}. For definiteness we give the various experimental bounds for $A_0/\tilde{m} =2$, and $\tilde m=\tilde m_0=2\mu=1$ TeV. Our results can be easily rescaled to other cases, by simply noticing that the bounds in the Table scale roughly as the ratio $\tilde m/({\textrm{1 TeV}})$ for $\Delta F=2$ observables and as $(\tilde m/{\textrm{1 TeV}})^2$ for $\Delta F=1$ ones. 
The scaling for different $\tilde m_0 / \tilde m$ and $A_0 / \tilde m$ can be read immediatly from (\ref{massinse}).

From Table \ref{table:susy} we see that flavor violation in the quark sector is again consistent with current bounds. The largest effects are predicted in $\Delta F=1$ processes mediated by dipole operators, and arise from observables such as $\epsilon'/\epsilon$ and $B\to X_s\gamma$. Analogously to the composite Higgs models of section~\ref{framework}, a rather robust prediction of this framework is that new effects in the neutron EDM should be around the corner. 

In the lepton sector we observe a significant improvement compared to Table \ref{table:generic}. This is because the Feynman diagrams of the most constraining processes ($\mu\rightarrow e\gamma$ and electron EDM) are now suppressed by $\alpha' / \alpha_s$ and $O(1)$ accidental numerical factors\footnote{Quantitatively $\alpha_s / \alpha' \sim 10$ and the loop function times the hypercharge couplings give another factor $\sim 3$, thus putting all together one finds an improvement by a factor $\sim 30$. Since the bound in Table \ref{table:generic} is satisfied with a suppression factor $\sim 200$, we are left with $(c_{12}^{u})_{LR}/(c_{11,12}^e)_{LR} \gtrsim 6$, which is tolerable at the level of this analysis. Notice that the tension disappears if the typical slepton mass is $2-3$ times larger than the typical squark mass.} with respect to those that generate the chromomagnetic operator used to fit $\Delta a_{CP}$. The model can be made fully consistent with data if a moderate suppression of $(c_{11,12}^e)_{LR}$ is arranged and/or if the slepton sector is taken to be slightly heavier than the squark sector.
A typical expectation is however that $\mu\rightarrow e\gamma$ and the electron EDM are the most promising observables in which new physics should manifest itself in the lepton sector.

\begin{table}[t]
\begin{center}
\begin{tabular}{c|c|c} \hline\hline
\rule{0pt}{1.2em}%
Coefficient & 
{Upper bound  } & Observables\cr
 \hline \hline
$(c^u_{12})_{LL}$   & $ 200 \, \left( \frac{1}{\epsilon^q_3} \right)^2$ &   $\Delta m_D$; $|q/p|, \phi_D$ \\
$(c^d_{12})_{LL}$   &  $  60 \, \left( \frac{1}{\epsilon^q_3} \right)^2 $ &   $\Delta m_K$; $\epsilon_K$; $\epsilon'/\epsilon$ \\
$(c^d_{13})_{LL}$   &  $ 20 \, \left( \frac{1}{\epsilon^q_3} \right)^2$ &  $\Delta m_{B_d}$; $S_{\psi K_S}$  \\
$(c^d_{23})_{LL}$   &  $ {10} \, \left( \frac{1}{\epsilon^q_3} \right)^2$ & $B\to X_s \gamma$  \\
\hline
$(c^u_{12})_{RR}$   & $ 2\times 10^3 \, \left( \frac{1}{\epsilon^u_3 } \right)^2$  &   $\Delta m_D$; $|q/p|, \phi_D$ \\
$(c^d_{12})_{RR}$   & $ 3\times 10^3 \, \left( \frac{1}{\epsilon^u_3 \, t_{\beta} } \right)^2$    &  $\Delta m_K$; $\epsilon_K$\\
$(c^d_{13})_{RR}$   & $ 8 \times 10^3 \, \left( \frac{1}{\epsilon^u_3 \, t_{\beta} } \right)^2$    &   $\Delta m_{B_d}$; $S_{\psi K_S}$ \\
$(c^d_{23})_{RR}$   & $ 2 \times 10^4 \, \left( \frac{1}{\epsilon^u_3 \, t_{\beta} } \right)^2$    &  $\Delta m_{B_s}$ \\
\hline
$\sqrt{(c^u_{12})_{LL} (c^u_{12})_{RR}} $ & ${60}  \,\,  g_{\rho} $  &    $\Delta m_D$; $|q/p|, \phi_D$ \\
$\sqrt{(c^d_{12})_{LL} (c^d_{12})_{RR}} $ & ${30}  \, \left( \frac{g_{\rho}}{ t_{\beta} } \right) $ &   $\Delta m_K$; $\epsilon_K$ \\
$\sqrt{(c^d_{13})_{LL} (c^d_{13})_{RR}} $ & ${100}  \, \left( \frac{g_{\rho}}{t_{\beta} } \right) $ &    $\Delta m_{B_d}$; $S_{\psi K_S}$ \\
$\sqrt{(c^d_{23})_{LL} (c^d_{23})_{RR}} $ & ${100}  \, \left( \frac{g_{\rho}}{ t_{\beta} } \right) $ &   $\Delta m_{B_s}$ \\
\hline
$(c^u_{12})_{LR}$   & ${90}   $  &  $\Delta m_D$; $|q/p|, \phi_D$ \\
$(c^u_{12})_{RL}$   & ${2\times 10^3}  $  &  $\Delta m_D$; $|q/p|, \phi_D$ \\
$(c^d_{12})_{LR}$   & {2} &$\epsilon'/\epsilon$  \\
$(c^d_{12})_{RL}$   & {2} &$\epsilon'/\epsilon$ \\
$(c^d_{13})_{LR}$   & ${2 \times 10^3}  $  &   $\Delta m_{B_d}$; $S_{\psi K_S}$  \\
$(c^d_{13})_{RL}$   & ${200} $  &   $\Delta m_{B_d}$; $S_{\psi K_S}$  \\
$(c^d_{23})_{LR}$   & $20  $  &  $B\to X_s \gamma$ \\
$(c^d_{23})_{RL}$   & $8  $  &  $B\to X_s \gamma$ \\
\hline
$(c^u_{11})_{LR}$   & $ {0.4}   $ &  EDMs  \\
$(c^d_{11})_{LR}$   & $ {0.09}   $  &  EDMs \\
\hline\hline 
$(c^e_{12})_{RR}$   & $ 4 \times 10^{4} \,   \left( \frac{g_\rho}{4\pi}\right) \left( \frac{\epsilon_3^e}{\epsilon_3^\ell t_\beta}\right)$ &  $\mu \rightarrow e\gamma $ \\
$(c^e_{12})_{LL}$   & $ 4 \times 10^{3} \,  \left( \frac{g_\rho}{4\pi}\right) \left( \frac{\epsilon_3^e}{\epsilon_3^\ell t_\beta}\right)$ &  $\mu \rightarrow e\gamma $ \\
$(c^e_{12})_{LR, RL}$   & $  0.6 $ &  $\mu \rightarrow e\gamma $ \\
$(c^e_{11})_{LR}$   & $ 0.5 $  &  electron EDM$ $ \\
\hline\hline 

\end{tabular}

\caption{\label{table:susy}\footnotesize{Upper bounds on the dimensionless coefficients of the mass insertions in the notation (\ref{eq:deltas:susy}), with $\tilde{m}=\tilde m_0=2\mu=1$ TeV and $A_0/\tilde{m}=2$. See text for details on the parameter scaling of the various bounds. We used the abbreviation $t_{\beta}=\tan\beta$. Note that $c\sim 4$ for the chromomagnetic $\Delta C=1$ operator. 
Notice that the combinations  $g_\rho \epsilon_3^q \epsilon_3^u$ and $g_\rho \epsilon_3^\ell \epsilon_3^e$ are fixed to be respectively $y_t^{(SM)}/\sin\beta$ and $y_\tau^{(SM)}/\cos\beta$.
The experimental constraints on the flavor-changing mass insertions in the quark sector are taken from \cite{Isidori:2010kg}. For the quark EDMs and the lepton sector see text. To minimize the constraints in the lepton sector we assumed~(\ref{R}). 
}
}

\end{center}
\end{table}

\clearpage

\subsection{R-Parity Violation}
\label{RPVsection}

In the absence of additional symmetries, the Lagrangian of the MSSM contains lepton- and baryon-number violating interactions already at the renormalizable level. From an effective field theory perspective it is important to ask whether these interactions are sufficiently suppressed in models of Partial Compositeness, or if some new dynamical assumption must be imposed to render these models phenomenologically viable. This is the purpose of the present section.

Following standard conventions, we write the (renormalizable) superpotential terms as:
\ba\label{RPVop}
W_{\not B}&=&\frac{1}{2}\lambda''_{ijk}u_id_jd_k,\\\no
W_{\not L}&=&\frac{1}{2}\lambda_{ijk}L_iL_je_k+\lambda'_{ijk}L_iQ_jd_k+\mu_iL_iH_u.
\ea
There are also analogous soft terms, and of course higher dimensional operators, but their phenomenological impact turns out to be subleading with respect to those in~(\ref{RPVop}), and will hence be neglected. 
Partial Compositeness predicts the following structure for the couplings of these operators:~\footnote{It is conceivable that the same physics responsible for solving the $\mu$-problem also implies that $\mu_i\propto\mu$.}
\ba\label{BV}
\lambda''_{ijk}\sim 2g_{\not\rm B}\epsilon_i^u\epsilon_j^d\epsilon_k^d~~~~~~~~\lambda_{ijk}\sim 2g_{\not\rm L}\epsilon_i^\ell\epsilon_j^\ell\epsilon_k^e~~~~~~~~~~\lambda'_{ijk}\sim g_{\not\rm L}\epsilon_i^\ell\epsilon_j^q\epsilon_k^d~~~~~~~~~~~\mu_i\sim \frac{g_{\not\rm L}}{g_\rho}\epsilon_i^\ell \mu,
\ea
where $g_{\not\rm B}$ and $g_{\not\rm L}$ are couplings controlling the strength of baryon and lepton number violation. In the absence of any symmetry we expect $g_{\not\rm L}\sim g_{\not\rm B}\sim g_\rho$.

The structure~(\ref{BV}) has been obtained applying the general principles discussed above, and in particular interpreting $L_i$'s as 
 \emph{elementary} fields, as opposed to $H_d$ which is \emph{composite}. However, from a more effective perspective, this distinction is not necessarily justified. Specifically, if the lepton number is violated, then there is a priori no distinction between $H_d$ and $L_i$, as both fields are charged under the same representation of the SM group. Let us elaborate on this issue a bit further. 

A collection of the bounds on the couplings~(\ref{BV}) is reported in~\cite{Barbier:2004ez}. These are generally estimated in a (super-) field basis $L'_\alpha\equiv(H'_d,L'_i)$ where the Yukawa couplings of the charged leptons as well as those of the quarks are diagonal, and where the ``physical" Higgs $H'_d$ is identified with the component of the four-vector $L'_\alpha$ that acquires a vacuum expectation value (or, equivalently, such that the primed sneutrinos have vanishing vacuum expectation value). Our aim is to show that in this perhaps more phenomenological field basis the couplings $\lambda_{ijk}$, $\lambda'_{ijk}$, $\lambda''_{ijk}$, and $\mu_i$ have the very same structure as shown in~(\ref{BV}), so that the constraints of~\cite{Barbier:2004ez} straightforwardly apply to our framework as well.

The crucial observation is that in the field basis adopted in~(\ref{RPVop}) the scalar, neutral components of $L_\alpha\equiv(H_d,L_i)$ will generally acquire vacuum expectation values that, in the absence of hierarchies among the soft SUSY masses, parametrically scale as:
\ba\label{vac}
\langle L_\alpha\rangle\sim \epsilon_\alpha v_d,
\ea
with $\epsilon^\ell_\alpha\equiv(\epsilon_{H_d}=1,\epsilon^\ell_i)$. This is so because all the tadpole couplings involving $L_i$ (and arising after electro-weak symmetry breaking) are proportional to $\epsilon^\ell_i$.~\footnote{This conclusion is invalidated if additional sources of flavor violation are introduced.} It then follows that the field basis $L'_\alpha$ defined by the condition $\langle L'_\alpha\rangle=(v_d,0)$ is related to $L_\alpha$ via a unitary transformation $L_{\alpha}=U_{\alpha\beta}L'_\beta$, with entries $U_{\alpha\beta}\sim$ min$(\epsilon^\ell_\alpha/\epsilon^\ell_\beta, \epsilon^\ell_\beta/\epsilon^\ell_\alpha)$:
\ba
H_d\sim H_d'+\epsilon^\ell_i L_i'~~~~~~~~~~~~~~L_i\sim\epsilon^\ell_iH_d'+{\rm min}\left(\frac{\epsilon^\ell_i}{\epsilon^\ell_j}, \frac{\epsilon^\ell_j}{\epsilon^\ell_i} \right)L_j'.
\ea
This result can be equivalently understood as a consequence of the spurion family symmetry carried by the $\epsilon^\ell_i$'s. Plugging the above redefinition in~(\ref{RPVop}) we find that the structure of $W_{\not B,\not L}$ in the ``physical" basis $L_\alpha'$ is \emph{identical}, up to subleading corrections, to that in the original, UV basis $L_\alpha$, see Eq.~(\ref{BV}).

Using~(\ref{masses}),~(\ref{eq:conditions:eps}), and~(\ref{massleptons}) we can thus write the coefficients of the operators~(\ref{RPVop}) in the $L'_\alpha$-basis as:
\begin{eqnarray}
\lambda''_{ijk} &\sim& 2 g_{\not\rm B}  \left(\frac{\epsilon_3^q}{\epsilon_i^q}\right)\left(\frac{\epsilon_3^q}{\epsilon_j^q}\right)\left(\frac{\epsilon_3^q}{\epsilon_k^q}\right)(\epsilon_3^u)^3\frac{m_i^u m_j^d m_k^d}{m_t^3}\tan^2\beta, \\
\lambda_{ijk} &\sim& 2\frac{g_{\not\rm L}}{g_\rho} \left(\frac{\epsilon_i^\ell}{\epsilon_3^\ell}\right)\left(\frac{\epsilon_j^\ell}{\epsilon_3^\ell}\right)\left(\frac{\epsilon_3^\ell}{\epsilon_k^\ell}\right)(\epsilon_3^\ell)\frac{m_k^\ell}{v\cos\beta}, \label{LV1} \\
\lambda'_{ijk}&\sim & \frac{g_{\not\rm L}}{g_\rho}  \left(\frac{\epsilon_i^\ell}{\epsilon_3^\ell}\right)\left(\frac{\epsilon_j^q}{\epsilon_3^q}\right)\left(\frac{\epsilon_3^q}{\epsilon_k^q}\right)(\epsilon_3^\ell)\frac{m_k^d}{v\cos\beta}, \label{LV2} \\
\mu_i &\sim&  \frac{g_{\not\rm L}}{g_\rho} \epsilon_i^\ell \mu, \label{LV3}
\end{eqnarray}
up to appropriate coefficients, with $O(1)$ entries according to NDA. 

We are now ready to study the viability of the scenario. 
The first thing we observe is that one cannot simultaneously allow lepton- as well as baryon- number violation because in this case the model would predict an unacceptably fast proton decay. Taking the bounds on $p\to\pi^0 \ell^+$ from~\cite{Barbier:2004ez} we find:
\ba
\epsilon_3^\ell (\epsilon_3^u)^3\left(\frac{g_{\not\rm B}g_{\not\rm L}}{g_\rho^2}\right)
\left(\frac{g_\rho}{4\pi}\right)\frac{\tan^2\beta}{\cos\beta}
&\lesssim&
10^{-15}\left(\frac{\tilde m}{1~{\rm TeV}}\right)^{2} \, .
\ea
We are thus led to consider B and L violation separately, i.e. $g_{\not\rm L}g_{\not\rm B}\ll g_\rho^2$.
The strongest bounds on $W_{\not B}$ can be summarized as follows:
\begin{equation}
(\epsilon_3^u)^3\left(\frac{g_{\not\rm B}}{4\pi}\right)\tan^2\beta\lesssim30\left(\frac{150~{\rm MeV}}{\tilde \Lambda}\right)^{5/2}\left(\frac{\tilde m}{1~{\rm TeV}}\right)^{5/2}~~~~~~(pp\to K^+K^+)
\end{equation}
\begin{equation}\label{protondecay}
(\epsilon_3^u)^3\left(\frac{g_{\not\rm B}}{4\pi}\right)\tan^2\beta\lesssim5\times10^{-8}\left(\frac{\tilde m}{1~{\rm TeV}}\right)^{2}\frac{m_{3/2}}{1~{\rm eV}}~~~~~~(p\to K^+\tilde G)
\end{equation}
where ${\tilde \Lambda}\sim 150$ MeV in the notation of~\cite{Goity:1994dq}.
The first bound arises from constraints on dinucleon decay $pp\to K^+K^+$~\cite{Goity:1994dq}, and can be satisfied for a maximally strong coupling $g_{\not\rm B}$ and $\tilde m\sim1$ TeV as long as $\epsilon_3^u\lesssim0.2\times(50/\tan\beta)^{2/3}$, which for large $\tan\beta$ means that $(t_L,b_L)$ must be almost entirely composite. A weaker constraint is given by neutron-antineutron oscillation, a process which is further suppressed by mass insertions compared to $pp\to K^+K^+$. On the other hand, the requirement that the rate for $p\to K^+\tilde G$ be below the current experimental sensitivity~\cite{Choi:1996nk} is much stronger, but is generally avoided in theories where the gravitino $\tilde G$ is sufficiently heavy. In gauge-mediated SUSY breaking models, this implies a lower bound on the messenger scale
\begin{equation}
M_{\rm mess}\sim g_X \frac{\alpha}{4\pi} \frac{m_{3/2} M_P}{\tilde m}\gtrsim 10^{13}\,{\textrm{ GeV}}\left(\frac{1\,{\textrm{TeV}}}{\tilde m}\right) g_X,
\end{equation}
where $g_X$ is the coupling parametrizing the amount of SUSY breaking mediated to the visible sector. Since $M_{\textrm{mess}}$ has to be smaller than $\sim 10^{15}\,\textrm{GeV}g_X$ to avoid reintroducing the flavor problem, this leaves only a narrow window to realize the model within gauge mediation. 

As already emphasized below Eq.~(\ref{R}), the parameters $\epsilon_i^\ell/\epsilon_3^\ell$ are in principle undetermined. To derive the constraints on the lepton number violating coefficients in~(\ref{LV1}), (\ref{LV2}) and~(\ref{LV3}) we work under the assumption of ``normal hierarchy" $\epsilon_1^\ell<\epsilon_2^\ell<\epsilon_3^\ell$. Here, by far the larger effect arises from the last term in $W_{\not L}$, which implies a mixing between neutrinos and neutralinos. Requiring $m_\nu<1$ eV gives:
\ba
\left(\frac{g_{\not\rm L}}{g_\rho}\right)^2 (\epsilon_3^\ell)^2\lesssim10^{-12}\left(\frac{\tilde m}{1~{\rm TeV}}\right)\frac{1}{\cos^2\beta}~~~~~~~~(m_\nu<1~{\rm eV}).
\ea
It is clear that some symmetry argument must be invoked in order to satisfy the above constraint.\footnote{Even if we imposed the UV condition $\mu_i=0$, a non-zero value would be generated by the $\mu$ term after the field rotation $L_\alpha\to L'_\alpha$.} The basis-dependent condition is in general written as $(\mu,\mu_i)\propto(\langle H_d\rangle, \langle \tilde\nu_i\rangle)$ and requires a highly nontrivial relation among soft SUSY breaking terms. 
Other constraints on $W_{\not L}$ pose no serious problems on the model. 

Finally, a significant constraint on the RPV interactions comes from requiring that they do not erase the baryon asymmetry of the universe that is present at temperatures above the electroweak phase transition $T_c\sim 100$ GeV. In fact if an interaction that violates B-L comes into equilibrium above $T_c$, then it will completely wash-out any preexisting asymmetry once combined with sphalerons \cite{Klinkhamer:1984di}\cite{Kuzmin:1985mm}, that violate B+L but preserve B-L.
To avoid the washout one has to impose that the rate of the B-violating interactions is smaller than the Hubble rate all the way down to $T=T_c$. For sfermion masses of order of the TeV, the resulting bound \cite{Barbier:2004ez} turns out to be:
\begin{equation} \label{eq:bound:baryo}
|\lambda_{ijk}|, |\lambda_{ijk}'|, |\lambda_{ijk}''| < 3 \times 10^{-7} \, .
\end{equation}
In the present framework this bound can be satisfied only in a corner of the parameter space, in the presence of a mild suppression of the baryon violating interactions in the strong sector (for example $g_{\not\rm B}\sim 0.1\ll g_{\rho}$, $\tan\beta\sim 1$ and $\epsilon^u_3\sim 0.2$). Anticipating the discussion of the subsequent Section, such a small value for the couplings would lead to displaced vertices or missing energy signals, unless the LSP is the $\tilde{t}_R$ or the $\tilde{c}_R$.
Notice that, even assuming a very large preexisting asymmetry as in \cite{Affleck:1984fy}, the bound (\ref{eq:bound:baryo}) is not relaxed by more than one order of magnitude, because the washout is exponential.
Alternatively, the bound could be escaped if the baryon asymmetry is generated below $T_c$ by some other mechanism or through the RPV interactions themselves. This last possibility has been discussed in the past by many authors, like \cite{Dimopoulos:1987rk}-\cite{Mollerach:1991mu} and others.
In general however one has to be careful because the baryon asymmetry cannot be generated at first order in the B-violating interaction, as shown by \cite{Nanopoulos:1979gx}. As a consequence one tends to obtain a too-small asymmetry after requiring that the reaction is out of equilibrium.

We conclude that SUSY models of Partial Compositeness can be phenomenologically viable if:
\ba
g_{\not\rm B}\sim g_\rho~~~~~~~~~~{\rm and}~~~~~~~~~~g_{\not\rm L}\ll g_\rho. 
\ea
The case with moderate $\tan\beta$ is favored if Baryon number is violated, which can only be allowed if the gravitino is heavier than $\sim1$ GeV. Lepton number violation is instead strongly disfavored. It might arise at very high mass scales and be connected with neutrino mass generation, in which case we would expect $g_{\not\rm L}$ to be suppressed by the seesaw scale. 
Similar conclusions follow if one uses minimal flavor violation as an alternative organizing principle, as discussed recently in~\cite{Csaki:2011ge}.

\subsubsection{Collider Phenomenology}

As opposed to the composite Higgs models of Section~\ref{framework}, flavorful SUSY theories capable of saturating Eq.~(\ref{eq:deltaAcp}) have a very rich collider phenomenology. In fact, the parameter space favored by Eq.~(\ref{eq:susy:assumptions}) is already severely constrained by missing energy searches at the LHC. Most of the current bounds can however be evaded in R-parity-violating (RPV) scenarios with $\lambda''\neq0$.  

A thorough discussion of the LHC phenomenology of the RPV framework goes beyond the scope of this work. Few points are nevertheless worth a short discussion. The Partial Compositeness ansatz leads to a hierarchy between the RPV couplings (evaluated for simplicity at 1~TeV):
\ba\label{see}
\lambda''_{ijk}
&\sim&\left( \frac{ g_{\not\rm B}}{4\pi}\right) \left(
\frac{\tan\beta}{3}\right)^2\left(\frac{\epsilon_3^u}{0.5}\right)^{3}(\equiv\lambda_0)\times
\left\{\begin{array}{cc}
          2.7\times 10^{-3}&   (tbs)\\
                    0.6\times 10^{-3}&   (tbd)\\
         1.7\times 10^{-4}&    (cbs)\\
                  0.5\times 10^{-4}&    (cbd)\\
         1.7\times 10^{-6}&    (ubs)\\
         0.4\times 10^{-6}&    (ubd)
\end{array}\right.
\ea

As discussed above, this hierarchy makes sure that the parameters involving the first generations
are small enough to avoid experimental bounds while the couplings of the third generation are
large enough to allow the LSP to decay without the appearance of missing energy. Similar results hold in the case of Minimal-Flavor-Violating RPV~\cite{Csaki:2011ge}.


Even in the absence of large missing energy signatures, the model can still be constrained by multilepton searches. In~\cite{Allanach:2012vj} it was suggested to use same-sign dilepton events arising from top-quark decay. There it was shown that, in a very minimal RPV scenario with just the gluino and the right handed top squark, $\sim1$/fb of data was sufficient to rule out gluino masses below $\sim550$ GeV. Using the full 2011 dataset \cite{CMS} one can naively rescale the results of~\cite{Allanach:2012vj} and find that the bound is now pushed to $\sim600$ GeV. 
We generically expect similar bounds on the gluino mass to hold in our scenario, just as a consequence of the dominance of the $\lambda_{tbs}$ coupling among the R-parity breaking interactions. However, as we discuss below, there can be important exceptions to this conclusion.

At the LHC both R-parity conserving interactions and R-parity breaking ones can be responsible for the appearance of top quarks in the decay chain of a gluino.
The typical instance of the first kind is the super-QCD decay $\tilde g\to t\tilde t^*(\bar t \tilde t)$. This kind of process is easily avoided by making the top-stop final state kinematically unaccessible ($m_{\tilde g} < m_t + m_{\tilde t}$) while keeping some other squark lighter than the gluino. 

As a concequence of the hierarchy $g_{SM}\gg \lambda_{RPV}$ where $g_{SM}$ is a typical (gauge) coupling in the MSSM, production of top quarks through RPV interactions are only relevant for the decays of the LSP\footnote{The LSP can in principle also decay to its superpartner and a gravitino with a width given approximately by:
\begin{equation}
\Gamma_{\tilde G}\sim \frac{m_{LSP}^5}{8\pi m_{\tilde G}^2M_P^2},
\end{equation} corresponding to a very long lifetime:
\begin{equation}
\tau_{LSP}= 3\cdot 10^{10}\,\textrm{m}\left(\frac{300 \,\textrm{GeV}}{m_{LSP}}\right)^5\left(\frac{m_{\tilde G}}{1\, \textrm{GeV}}\right)^2.
\end{equation} Notice that for fixed $LSP$ mass this is a lower bound due to the constraint \eqref{protondecay} on proton decay.}. 
Let us analyze some of the possibilities.
The case of a neutralino LSP typically leads to the prompt decay to a $tbs$ final state, unless $m_{\chi}<m_t$. In this case the neutralino can easily lead to a displaced vertex:
\begin{equation}
\tau_{\chi-LSP}=0.1\,\textrm{mm}\frac{\beta}{\sqrt{1-\beta^2}} \left(\frac{150\,\textrm{GeV}}{m_{LSP}}\right)^5\left(\frac{\tilde m}{500\,\textrm{GeV}}\right)^4\lambda_0^{-2},
\end{equation}
where $\beta$ is the neutralino velocity in the lab frame and $\tilde m$ the typical squark mass. Similar considerations apply to the case of a chargino LSP.

Sleptons and left-handed squarks LSP are peculiar since these particles are not directly involved in the RPV superpotential. A slepton/sneutrino LSP goes through a 4 body decay to $\ell/\nu tbs$, which generically contains hard isolated leptons in addition to those coming from the top. The expected lifetime (assuming the final with state top to be kinematically allowed) is given by:
\begin{equation}
\tau_{\tilde \ell-LSP}=0.02\,\textrm{mm}\frac{\beta}{\sqrt{1-\beta^2}} \left(\frac{300\,\textrm{GeV}}{m_{LSP}}\right)^7\left(\frac{\tilde m}{500\,\textrm{GeV}}\right)^6\lambda_0^{-2},
\end{equation}
where we assumed a common mass $\tilde m$ for the s-particles. The decay may lead to a displaced vertex.
A left-handed squark has two possibilities: either a two body decay which can occur thanks to the small mixing with its right-handed partner or a four-body decay through an off-shell gluino (or gaugino in general). Both will give rise to top quarks. The former channel typically dominates for bottom squarks, leading to a prompt decay:
\begin{equation}
\tau_{\tilde b_L-LSP}=0.03\,\mu\textrm{m}\frac{\beta}{\sqrt{1-\beta^2}} \left(\frac{400\,\textrm{GeV}}{m_{LSP}}\right)\left(\frac{7.5\cdot 10^{-3}}{\theta^{\tilde b}_{LR}}\right)^2\lambda_0^{-2},
\end{equation}
where $\theta^{\tilde b}_{LR}\sim m_b A_0/\tilde m^2$ is left-right sbottom mixing angle.
The latter channel above mostly occurs for the first two generations squarks implying lifetimes of order:
\begin{equation}
\tau_{\tilde q_L-LSP}=4\,\mu\textrm{m}\frac{\beta}{\sqrt{1-\beta^2}} \left(\frac{400\,\textrm{GeV}}{m_{LSP}}\right)^7\left(\frac{\tilde m}{1\,\textrm{TeV}}\right)^6\lambda_0^{-2},
\end{equation}
where we assumed gluino exchange and a similar mass $\tilde m$ for all the superpartners.

Right-handed down squarks decay promptly to a two body final state involving tops. A remaining possibility is to consider as the LSP a right-handed up squark $\tilde{u}_i$, $i=1,2$ of the first two generations. Due to the form of the RPV superpotential its decay will not involve top quarks and will lead promptly to a $bs$ final state through the $\lambda_{ibs}''$ coupling\footnote{The case of a right-handed stop LSP is less straightforward and deserves a separate discussion. One needs to impose $m_{\tilde g}<m_t+m_{\tilde t}$ to avoid the tops from gluino decay. At this point the leading two-body decay of the gluino is to $\bar c\tilde t(c\tilde t^*)$ through RR mixing. This channel competes with the three-body $\tilde g\to W^-b\tilde t^*(W^-\bar b\tilde t)$ through an off-shell top quark. In order to avoid same-sign di-leptons this latter channel must be subdominant. Parametrically this happens for $g_\rho\sim 1$ when the $\epsilon_3^u$ is maximal.}.

\begin{figure}[t]
\begin{center}
\includegraphics[width=0.45\textwidth]{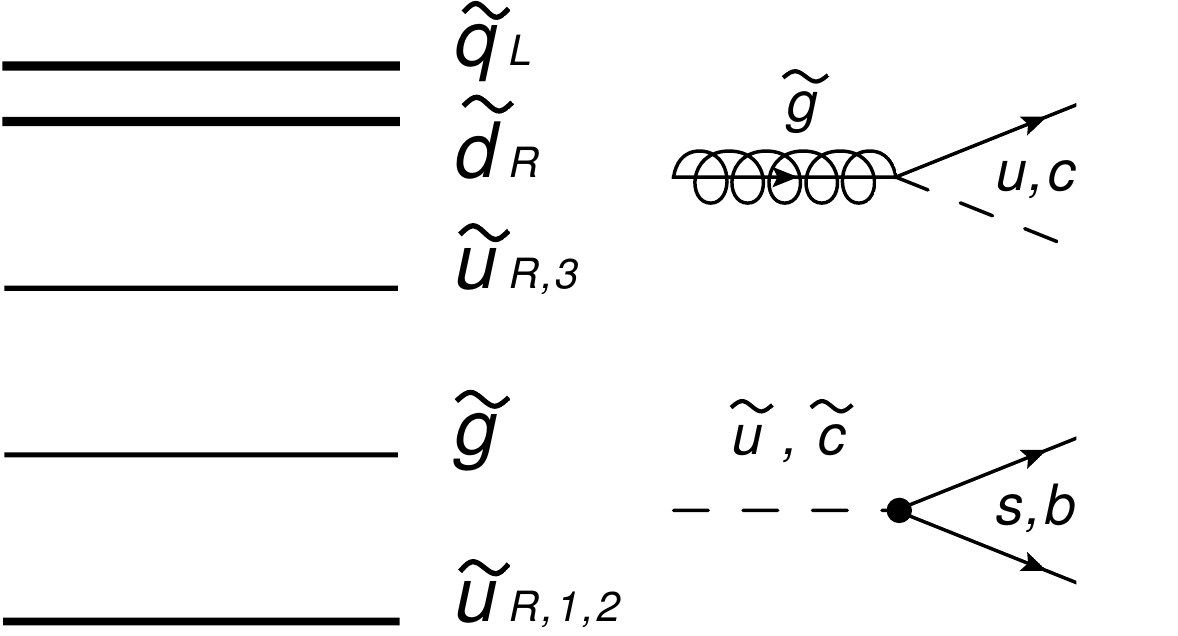}
\caption{\footnotesize{A prototypical spectrum with R-handed up squark LSP avoiding missing energy signals and isolated lepton events. See text for details.}}
\label{fig:RPV}
\end{center}
\end{figure}

Following these considerations, a prototypical example of a supersymmetric spectrum evading missing energy constraints as well as lepton searches is shown in Figure \ref{fig:RPV}. The only squarks below the gluino are $\tilde u_R$ and $\tilde c_R$, one of which is also the LSP. The sleptons, the left-handed up-squarks, as well as the down-type squarks have masses $\gtrsim m_{\tilde g}$. The neutralinos/charginos can be heavier or lighter than $\tilde g$. 
With this configuration the gluino will be pair-produced and subsequently decay to a quark-squark pair $u_i\tilde u_i$ where $i=1,2$, with the squark decaying promptly to a $bs$ final state with a lifetime of order:
\ba
\tau_{u_R(c_R)-LSP}=4\,\mu{\textrm m}(0.3\, {\textrm{nm}})\frac{\beta}{\sqrt{1-\beta^2}} \left(\frac{400\,\textrm{GeV}}{m_{LSP}}\right)\lambda_0^{-2}.
\ea
The typical event is thus an experimentally challenging 6 jet final state, with no missing energy nor displaced vertices. 
At present the searches for three-jet resonances set a lower bound of 460 GeV on the gluino mass assuming unit branching fraction \cite{CMS:EXO6j}.
%
It is worth noticing that a spectrum of the kind shown in Figure~\ref{fig:RPV} follows from Partial Compositeness if the flavor-diagonal masses of $\tilde m_{q,d,\ell,e}$ are somewhat larger than or of the order of the gluino mass, while the up-squark families are split such that 
\ba
m_{\tilde{u}_R,\tilde c_R} < m_{\tilde{g}} \lesssim m_{\tilde{t}_R}. 
\ea
This typically requires a mild suppression of $\tilde m_u$ compared to $\tilde m\sim\tilde m_{q,d,\ell,e}$ and $g_\rho\sim 1$.

Quantitatively, the suppression in the number of top events can be roughly estimated to be the ratio between a 2-body and a 3-body decay, i.e. a factor of order $10^2$. Notice also that the flavor-violating decay of the gluino into a top and an up-type squark of the second generation, which is kinematically allowed, is suppressed by about the same amount.
In particular, since the production cross section scales approximately as the sixth power of the inverse mass of quarks and gluinos, we expect that in such configurations the bound on $m_{\tilde{g}}$, as estimated in \cite{Allanach:2012vj}, is reduced by about a factor~2.

It is clear that, even assuming the picture of Figure \ref{fig:RPV}, the squarks and the gluino cannot be arbitrarily light. A simple and robust lower bound is obtained considering the LSP pair production and its decay into 2 jets
\begin{equation}
pp\to \tilde u_i\tilde u_i^*, \tilde u_i\tilde u_i, \tilde u_i^*\tilde u_i^*\to 4j,
\end{equation}
which is constrained by a recent CMS analysis \cite{cmsdijet} looking for pair produced dijet resonances. We can do a `back of the envelope' estimate using their limits. We assume a 2\% acceptance, we calculate the cross sections using \textsc{MadGraph5} \cite{Alwall:2011uj}, and we include a K-factor of 1.5. Putting for simplicity $m_{\tilde u}=m_{\tilde g}$ we find a lower bound of 400\,GeV for a right-handed up squark LSP and 350\,GeV for a right-handed charm squark LSP. The difference between these numbers is due to the pdf enhancement of the process $pp\to \tilde u_R \tilde u_R$ due to t-channel gluino exchange in the case of the up squark.

\section{Conclusions} 
\label{sec:conclusions}

Partial Compositeness offers an elegant solution to the SM flavor puzzle, and serves as a powerful organizing principle for flavor-violation in theories beyond the SM. In this paper we discussed its implementation within two of the most compelling models for the weak scale: Composite Higgs and weak-scale Supersymmetry with and without R-parity. 
We studied the compatibility of these scenarios with the large CP asymmetry $\Delta a_{CP}$ in D meson decays recently measured by LHCb~\cite{Aaij:2011in} and CDF~\cite{CDFupdate}. Our results are summarized in Table \ref{table:generic} and Table \ref{table:susy} for two illustrative benchmark points. 


Flavor-violation in Composite Higgs models is dominantly mediated by the dimension-6 operators in Eqs.~(\ref{eq:gen:1})-(\ref{eq:gen:2}), and the most phenomenologically favorable scenarios are those in which the flavor sector is maximally strong (i.e. those with $g_\rho\sim4\pi$).
In this regime the relative size of the coefficients of the $\Delta F=2$ operators compared with the $\Delta F=1$ ones is dictated by the flavor structure only, which implies a double suppression of the former class.
This makes it possible to generate the asymmetry in the charm sector without conflict with the strong bounds on $\Delta F=2$ observables.
An asymmetry $\Delta a_{CP}$ at the percent level can in fact be accommodated provided the mass scale of the resonances of the new, strong flavor sector is not far from $m_\rho\sim10$ TeV. 

If this framework is responsible for saturating the observed value of $\Delta a_{CP}$, then new physics contributions close to the current bounds are expected in the neutron EDM as well as in $\epsilon'/\epsilon$, $B\to X_s\gamma$ and $\epsilon_K$.
In principle the effect in $\epsilon_K$ may be welcome, since it can improve the CKM fit. This however totally depends on the precise value of the $O(1)$ coefficient of the corresponding operator, on which we did not make any precise assumption. Moreover the large down-quark contribution to the neutron EDM suggests that, in a realistic model, the operators involving down quarks need to be moderatly suppressed with respect to the ones involving up-type quarks. We thus believe that a complete model is necessary in order to asses the viability of improving the CKM fit in this framework.

In addition, new physics effects in $K^+\to\pi^+\bar\nu\nu$ are predicted to be within the reach of the NA62 experiment, see (\ref{eq:KtoPiNuNu}). Among these effects, however, the large contribution to the neutron EDM from the chromoelectric dipole operator made of up quarks represents the most robust signature of the model, since this operator has the same structure of the one used to fit the LHCb result~(\ref{eq:deltaAcp}), and hence is expected to have a numerical coefficient of comparable magnitude. Unless additional fields are introduced at the TeV scale, a Higgs boson with a mass $\sim125$ GeV will require a fine-tuning of the order of percent to permille, and the above signatures will be the only ones within the reach of the LHC.

Some additional dynamical assumptions have to be made to render Composite Higgs models of Partial Compositeness compatible with the stringent bounds from the lepton sector. On the other hand, much more freedom characterizes the Supersymmetric scenario,  where such constraints can be easily evaded as long as the typical slepton mass is comparable or slightly bigger than the typical squark mass.
To be specific, slepton masses larger than quark masses by a factor of $2-3$ are enough to satisfy the experimental bound even without assuming the A-terms to be larger in the quark sector.
We argued that Supersymmetric models of Partial Compositeness realize the `disoriented A-terms' scenario advocated in~\cite{Giudice:2012qq}, and therefore provide an ideal framework to explain the LHCb result. A robust implication  is again the presence of sizable corrections to the neutron EDM induced by the chromomagnetic dipole operators, but now possibly together with new effects in $\mu \rightarrow e \gamma$ and the electron EDM.

In order to saturate the measured $\Delta a_{CP}$, Supersymmetric models must approximately satisfy Eq.~(\ref{eq:susy:assumptions}). This is generally accomplished with relatively large A-terms and soft masses below the TeV scale. Large A-terms tend to enhance the radiative stop corrections to the Higgs boson mass, and are therefore favored by the recent LHC data~\cite{Higgs}.

Sparticle masses below the TeV scale are allowed if R-parity is violated. 
We find that a sufficient condition to evade the most stringent collider constraints, currently arising from searches involving missing energy and isolated leptons, is to switch on baryonic R-parity violation, taking either $\tilde u_R$ or $\tilde c_R$ to be the LSP and all the other scalar superpartners heavier than the gluino.
We showed that the renormalizable superpotential term $u_id_jd_k$, with coefficients at their natural values, is compatible with data provided the gravitino is sufficiently heavy to forbid $p\to K^+\tilde G$. On the other hand, some symmetry has to be invoked in order to suppress lepton number violation. 

We remark that in the Supersymmetric case the above scenario is fully within the reach of the LHC, since superpartners at the TeV scale are required.
On the contrary the Composite Higgs case suggests new physics at about 10 TeV, making it difficult to detect new particles unless the model is complicated in such a way that some of them are made light without disturbing flavor.

Finally we quote the proposal \cite{Isidori:2012boh} about the crucial point of checking whether the large CP asymmetry $\Delta a_{CP}$ in D meson decays comes from new physics or can instead be explained within the SM. In fact if, as we assumed, a chromomagnetic dipole operator is responsible for the enhancement of the CP asymmetry in D decays into $P^+ P^-$ ($P=\pi,K$), then the corresponding electromagnetic dipole will also be present, giving a large contribution to the CP asymmetry in \emph{radiative} D decays into $P^+ P^- \gamma$, especially with $M_{PP}$ close to the $\rho$ or the $\phi$ peak.
The observation of such asymmetries at the level of several percent would be a striking indication of physics beyond the SM in the dipole operators, thus corroborating the picture we discussed in this paper.

\section*{Acknowledgments}

We thank Kaustubh Agashe, Giancarlo D'Ambrosio, Gian Giudice, Yuval Grossman, Ulrich Haisch, Gino Isidori, Rakhi Mahbubani, Gilad Perez, Michele Redi, Luca Silvestrini, Raman Sundrum, Andreas Weiler and Andrea Wulzer for useful discussions.
The research of BK, PL, DP, and RR is supported by the Swiss National Science Foundation under grants 200020-138131 and 200021-125237. The work of LV is supported by the DOE Office of Science and the LANL LDRD program.

\section*{Appendix}
\appendix


\section{Standard Model inputs}
\label{app:formulae}

In our numerical estimates we take $\alpha_{em}=e^2/4\pi=1/137$, the Cabibbo angle $\lambda=0.22$, the Higgs vev $v=174$ GeV, and employ the following value of $\alpha_s$ at $1$~TeV \cite{PDG}:
\begin{equation}
\alpha_s(1~\textrm{TeV})=0.089
\end{equation}
In addition, the quark masses evaluated at 1~TeV are given in Table \ref{table:quarkmasses}, see~\cite{Xing:2007fb}.
The strong gauge coupling and running quark masses at different renormalization scales are obtained through the standard 1-loop RG evolution in the SM. The sine of the electroweak angle, $\sin\theta_w$, and the elements of the CKM matrix, $V_{ij}$, are taken from the Particle Data Group~\cite{PDG}.

\begin{table}[h]
\begin{center}
\begin{tabular}{cccccc}
$m_u$&$m_d$&$m_s$&$m_c$&$m_b$&$m_t$\\
\hline
$1.10_{-0.37}^{+0.43}$& $2.50_{-1.03}^{+1.08}$&$47_{-13}^{+14}$&$0.53\pm0.07$&$2.43\pm0.08$&$150.7\pm 3.4$
\end{tabular}
\caption{\label{table:quarkmasses}  \footnotesize{ Running quark masses  at $\mu=1~\textrm{TeV}$ (in MeV for $u,d,s$; in GeV for $c,b,t$).}}
\end{center}
\end{table}

\section{Phenomenological Constraints in the Composite case}
\label{pheno}

We are interested in constraining the coefficients $c_{ij, {\rm SM}}^{ab}$, $c_{ij}^{ab}$, and $c_{ijkl}^{abcd}$ in Eqs. (\ref{eq:gen:1})--(\ref{eq:gen:2}) evaluated at the scale $\sim m_\rho$. Since the strongly coupled regime $g_\rho\sim4\pi$ tends to be favored, in estimating the bounds we take the effect of the RG evolution from scales of order $m_\rho\sim\Lambda\sim10$ TeV down to the appropriate hadronic scale.

\subsection{Dipole Operators}

The first transitions we consider are those mediated by~(\ref{eq:gen:1}). The most stringent constraints on this class of operators come from $B\to X_s$ transitions, $\epsilon'/\epsilon$, and the neutron EDM $d_n$. 

For what concerns $b\to s\gamma$, we follow~\cite{Agashe:2008uz}\cite{Gedalia:2009ws} and define the following effective Hamiltonian:
\begin{equation}
\frac{G_F}{\sqrt{2}}V_{tb}V_{ts}^*\frac{m_b}{4\pi^2} \left[C_{7\gamma}e\overline{s_L}\sigma^{\mu\nu}b_R F_{\mu\nu}+C'_{7\gamma}e\overline{s_R}\sigma^{\mu\nu}b_L F_{\mu\nu}\right],
\end{equation}
where, at the matching scale, we find:
\begin{equation}
C_{7\gamma}(m_\rho)=c_{23,\gamma}^{qd}~\frac{4\pi^2\sqrt{2}}{G_F\Lambda^2}\frac{\lambda^2}{V_{tb}V_{ts}^*}.~~~~~~~~~~~~C'_{7\gamma}(m_\rho)=c_{23,\gamma}^{dq}~\left(\frac{4\pi^2\sqrt{2}}{G_F\Lambda^2}\frac{\lambda^2}{V_{tb}V_{ts}^*}\right)\frac{m_s(m_\rho)}{m_b(m_\rho)}\frac{1}{\lambda^4}.
\end{equation}

We take the bounds from~\cite{Altmannshofer:2011gn}, which approximately read:
\ba\label{btos}
|C'_{7\gamma}(m_b)|\lesssim0.2~~~~~~~~~~~|{\rm Re}C_{7\gamma}(m_b)|\lesssim0.06~~~~~~~~~~~|{\rm Im}C_{7\gamma}(m_b)|\lesssim0.2.
\ea
The RG evolution can be taken into account by identifying $C_7^{(')}(\mu)\approx C_7^{(')}(m_\rho)\times\eta^{16/3\beta_0}$. With these assumptions, eq.~(\ref{btos}) implies the constraints on $c_{23}^{qd}$, $c_{23}^{dq}$ shown in Table \ref{table:generic}. It should be noted that the bounds~(\ref{btos}) are somewhat conservative because in deriving them we assumed that only one operator at a time is turned on. When more operators are active the bounds turn out to be somewhat weaker, as shown in Fig. 6 of~\cite{Altmannshofer:2011gn}.

The relevance of the observable $\epsilon'/\epsilon$ in these scenarios was first pointed out in~\cite{Gedalia:2009ws}. Using the notation of~\cite{Gedalia:2009ws}, the coefficients of the operators $Q_G=H^\dagger\overline{s_R}\sigma^{\mu\nu}g_sG_{\mu\nu}d_L$ and $Q'_G=H\overline{s_L}\sigma^{\mu\nu}g_sG_{\mu\nu}d_R$ read
 \ba
 C_G=c_{21,g}^{dq}~\frac{m_s\lambda}{\Lambda^2}~~~~~~~~~~~~~~~~ C'_G=c_{21,g}^{dq}~\left(\frac{m_s\lambda}{\Lambda^2}\right)\frac{m_d}{m_s\lambda^2},
 \ea
and must satisfy~\cite{Gedalia:2009ws}~\footnote{In \cite{Gedalia:2009ws} the running is erroneously done as if the quark mass were not included in the definition of the operator. We thank Gino Isidori for discussions on this point.}:
 \ba\label{run}
 \left|\frac{{\rm Im}(C^{(')}_G(1~{\rm GeV}))}{m_s(1~{\rm GeV})/v}\right|\approx\left|\frac{{\rm Im}(C^{(')}_G(m_\rho))}{m_s(m_\rho)/v}\right|\times \eta^{14/3\beta_0} \lesssim\frac{1}{(58~{\rm TeV})^2}.
 \ea

Next we consider the neutron EDM $d_n$. 
Assuming that some mechanism is invoked to suppress the $\theta$ term, then the largest contributions to $d_n$ from new physics above the TeV scale arise from the flavor-diagonal Lagrangian:
\ba
-i\frac{d_q}{2}\overline{q}\sigma^{\mu\nu} F_{\mu\nu}\gamma^5q-i\frac{\tilde d_q}{2}\overline{q}\sigma^{\mu\nu} g_sG_{\mu\nu}\gamma^5 q.
\ea
The determination of the actual relation between the parton-level parameters $d_q,\tilde d_q$ and $d_n$ suffers from large QCD uncertainties and it is fair to say that, to date, there exists no completely reliable estimate. A full lattice simulation of $d_n(d_q,\tilde d_q)$ is currently lacking, and one is therefore forced to rely on semi-analytic, and hence necessarily ambiguous methods. From NDA we expect $d_n=\sum_q (a_qd_q+e\tilde a_q\tilde d_q)$, with $a_q,\tilde a_q$ unspecified numbers $O(1)$. The non-relativistic quark model consistently predicts $a_d=-4a_u=4/3$, but unfortunately gives no information about $\tilde a_q$. An alternative method makes use of QCD sum rules. Employing this latter option Pospelov and Ritz find~\cite{Pospelov:2000bw}:
\ba\label{nEDM}
d_n=(1\pm0.5)\left[1.4(d_d-0.25d_u)+1.1e(\tilde d_d+0.5\tilde d_u)\right].
\ea
This result is in good agreement with the first two methods, but perhaps underestimates the systematic uncertainty. Recently, it was shown in~\cite{Hisano:2012sc} that an identical analytical method but different numerical inputs~\footnote{The neutron wavefunction in~\cite{Hisano:2012sc} is extracted by recent lattice data and turns out to be roughly a factor 2 bigger than the semi-analytic estimate adopted by~\cite{Pospelov:2000bw}.} lead to a suppression of a factor $\sim 4$ compared to~(\ref{nEDM}). Rather than dwelling on the significance of these discrepancies, we decide to adopt the central value of the conservative formula~(\ref{nEDM}), especially in view of its compatibility with the NDA estimate. The reader should be warned that with the present knowledge this is however an arbitrary choice.

For the coefficients at the scale $m_\rho$ from~(\ref{NDA}) we have:
\ba
d_q(m_\rho)\sim2\frac{m_q(m_\rho)}{\Lambda^2}e~~~~~~~~~~~~~\tilde d_q(m_\rho)\sim2\frac{m_q(m_\rho)}{\Lambda^2}.
\ea
After having renormalized the coefficient at the appropriate scale $\sim1$ GeV according to $d_q(\mu)=\eta^{4/3\beta_0}d_q(m_\rho)$ and $\tilde d_q(\mu)=\eta^{2/3\beta_0}\tilde d_q(m_\rho)$, we impose the bound $d_n\lesssim2.9\times10^{-26}$cm e~\cite{PDG}. In deriving the results of Table \ref{table:generic} we did not allow for cancellations among the different contributions in~(\ref{nEDM}).

\subsection{Penguin Operators}

Under the assumption that the Higgs doublet is part of the strong sector, the dominant contributions to $\Delta F=1$ processes are mediated by dipole operators as well as the penguin operators arising from~(\ref{eq:gen:1bis}). The dipole operators have been discussed above, here we focus on the latter. 

After electro-weak symmetry breaking, Eq.~(\ref{eq:gen:1bis}) leads to a correction:
\ba\label{correction}
\delta J_\mu^{(Z)}=2{\epsilon_i^{a} \epsilon_j^{b} g_{\rho}^2 }\frac{(4\pi)^2}{g_{\rho}^2}\, \frac{v^2}{\Lambda^2} c_{ij}^{ab} \overline{f}^{a}_i \gamma^{\mu} f^{b}_j
\ea
of the SM $Z^0$ boson current:
\ba
J_\mu^{(Z)}=\frac{1}{2}(-1+4s_W^2)\bar \ell \gamma^\mu \ell+\frac{1}{2}\bar\ell\gamma^\mu\gamma^5 \ell+\bar\nu_L\gamma^\mu \nu_L+\dots.
\ea
Integrating out the $Z^0$ boson, the powers of the electro-weak vacuum and gauge couplings cancel against the $Z^0$ mass, and one obtains the effective Lagrangian:
\ba\label{penguin}
\epsilon_i^{a} \epsilon_j^{b} \, \frac{(4\pi)^2}{\Lambda^2} c_{ij}^{ab} ~\overline{f}^{a}_i \gamma^{\mu} f^{b}_j~J_\mu^{(Z)}.
\ea
We note that $\Delta F=2$ operators obtained by sandwiching two powers of~(\ref{correction}) are suppressed by a factor of order $(4\pi)^2v^2/\Lambda^2\sim5\%$ compared to the genuinely UV contributions of~(\ref{eq:gen:2}) (see next subsection), and will hence be ignored.

As shown in Table~\ref{table:generic}, the most important constraints on~(\ref{penguin}) and~(\ref{correction}) -- and therefore indirectly on~(\ref{eq:gen:1bis}) -- come from $\mu\to e$ conversions and $Z^0\to\overline{b}b$ measurements at LEP. We will come back to these in subsequent Sections. Here we notice that additional constraints on~(\ref{penguin}) can be obtained from the relatively clean processes $B\to X_s\ell^+\ell^-$, $B_s\to\ell^+\ell^-$, and $K^+\to\pi^+\nu\nu$. We specialize to $B\to X_s\ell^+\ell^-$, since it provides the most stringent bound among the semi-leptonic meson decays just mentioned, even taking into account the recent improvements in $B_s\to \mu^+\mu^-$~\cite{LHCbLaThuile}.

To this end we derive the Hamiltonian~\cite{Altmannshofer:2011gn}:
\ba
\frac{G_F}{\sqrt{2}}V_{tb}V^*_{ts}\frac{e^2}{4\pi^2}\left[C_{10}\bar s_L\gamma^\mu b_L\ell\gamma^\mu\gamma_5\ell+C'_{10}\bar s_R\gamma^\mu b_R\ell\gamma^\mu\gamma_5\ell\right],
\ea
where:
\ba
C_{10}=\frac{2\pi^2\sqrt{2}}{G_F e^2}\frac{\lambda^2}{V_{tb}V^*_{ts}}\frac{(4\pi)^2}{\Lambda^2}c_{23}^{qq}(\epsilon_3^q)^2~~~~~~~~~~~~~~C'_{10}=\frac{2\pi^2\sqrt{2}}{G_F e^2}\frac{\lambda^2}{V_{tb}V^*_{ts}}\frac{(4\pi)^2}{\Lambda^2}c_{23}^{dd}(\epsilon_3^u)^2\frac{m_sm_b}{\lambda^4m_t^2}.
\ea
The coefficients of the operators with $\ell\gamma^\mu\gamma_5\ell\to\ell\gamma^\mu\ell$ are suppressed compared to the ones above by $1-4\sin^2\theta_w=0.08$, and have comparable experimental bounds, hence they will be neglected. The QCD renormalization down to lower energies vanishes, so the above values approximately hold at the hadronic scale as well. The bounds can be read from Fig.2 of~\cite{Altmannshofer:2011gn},
\ba
|{\rm Re}(C_{10})|\lesssim1.5~~~~|{\rm Im}(C_{10})|\lesssim2.5~~~~~~~~~~~|{\rm Re}(C'_{10})|\lesssim4~~~~|{\rm Im}(C'_{10})|\lesssim3,
\ea
and lead to the results of the Table. 

As a final remark we notice that the NA62 experiment at CERN \cite{NA62} is expected to measure the branching fraction of the process $K^+\to\pi^+\bar\nu\nu$ at the 10\% level after 2 years of running. This represents more than one order of magnitude improvement on the present measurement \cite{PDG}. Using eq.~\eqref{penguin} we find that an observable signal is possible in the non supersymmetric case if:
\begin{equation} \label{eq:KtoPiNuNu}
c_{12/21}^{qq}\gtrsim 1.8~(\epsilon_3^{u})^2\left(\frac{g_\rho}{4\pi}\right)^2~~~~~~~~~~~~~~~~c_{12/21}^{dd}\gtrsim 90~(\epsilon_3^{q})^2\left(\frac{g_\rho}{4\pi}\right)^2.
\end{equation}
In the supersymmetric case we expect the new physics contribution to be suppressed at least by a factor $\alpha_w / \alpha_s$, because the dominant effect comes from diagrams involving charginos \cite{Buras:1997ij}. Notice however that the effect can be larger if $g_\rho$ is smaller than $4\pi$.

\subsection{$\Delta F=2$ Transitions}

We now briefly comment on the operators in~(\ref{eq:gen:2}).
The Wilson coefficient of the operator $(\bar s_R\, d_L)(\bar s_L d_R)$ is strongly constrained by $K-\overline{K}$ mixing data, and represents one of the most stringent bounds on this scenario. We will not reproduce the derivation of the bound here because it is somewhat a standard result  (see for example~\cite{Csaki:2008zd}), but notice that in order to minimize its impact one should assume a maximally strong sector, $g_\rho\sim4\pi$. Even so, for $\Lambda\sim10$ TeV a mild suppression of the coefficient $c_{dqqd}^{2121}$ is preferable. 

As can be seen from Table~\ref{table:generic}, the next to sensitive $\Delta F=2$ process is $B_d$ oscillation via the operator $(\bar b_L \gamma^\mu d_L )^2$. Despite the larger RG enhancement obtained by $(\bar b_R\, d_L)(\bar b_L d_R)$, the Wilson coefficient of the former is predicted by the present framework to be much larger at the matching scale $\sim m_\rho$ in the regime $\epsilon_3^q\gtrsim\epsilon_3^u$. Intriguingly, possibly large effects in this channel seem to be required to fit the current data on the unitarity triangle. We will not investigate this issue any further.

\subsection{Top flavor changing neutral currents} 
\label{sec:top}
The set of leading gauge invariant operators contributing to neutral flavor-changing two-body decays of the top quark are the following\footnote{Those operators contributing only to $t\to q h$ are not listed as they are not present if, as remarked in Section 3, a mechanism is assumed to align the Higgs boson couplings to fermions with the Yukawas.}:
\begin{subequations}
  \begin{alignat}{2}
&\mathcal O^t_{H1}=\bar q_{iL}\gamma^\mu q_{3L}\, iH^\dagger  \overleftrightarrow D_\mu H\qquad&& \mathcal O^t_{H2}=\bar q_{iL}\gamma^\mu \tau^aq_{3L}\, i H^\dagger \tau^a  \overleftrightarrow D_\mu H\\
&\mathcal O^t_{H3}=\bar u_{iR}\gamma^\mu u_{3R}\, iH^\dagger  \overleftrightarrow D_\mu H\\
&\mathcal O^t_{F1}=\bar u_{iR}\sigma^{\mu\nu} H^T q_{3L}\, e F_{\mu\nu}\qquad&& \mathcal O^t_{F2}=\bar{q}_{iL}H^*\sigma^{\mu\nu} u_{3R}\, e F_{\mu\nu}\\
&\mathcal O^t_{G1}=\bar u_{iR}\sigma^{\mu\nu} T^A H^T q_{3L}\, g_s G^A_{\mu\nu}\qquad&& \mathcal O^t_{G2}=\bar{q}_{iL}H^*\sigma^{\mu\nu} T^Au_{3R}\, g_s G^A_{\mu\nu}.
 \end{alignat}
\end{subequations}
Expanding the operators in terms of the physical fields we have that:
\ba\label{lagtopFCNC}
\sum_{I} \frac{c_I}{m_\rho^2} \mathcal O^t_I&\equiv& \frac{e}{2m_t} F_{\mu\nu}\bar u_i(g_\gamma^{iL}P_L+g_\gamma^{iR}P_R)\sigma^{\mu\nu} t+\frac{g_s}{2m_t} G^A_{\mu\nu}\bar u_i(g_g^{iL}P_L+g_g^{iR}P_R)\sigma^{\mu\nu} T^A t\\\no
&+&\frac{g}{2 \cos\theta_w}  Z_\mu\,\bar u_i(g_Z^{iL}P_L+g_Z^{iR}P_R)\gamma^\mu t + {\rm h.c.},
\ea
where, for any given flavor (up or charm) of the outgoing quark,
\begin{equation}
g_Z^L=(-c_{H1}+c_{H2})\frac{2v^2}{m_\rho^2},\qquad g_Z^R=-c_{H3}\frac{2v^2}{m_\rho^2},
\end{equation}
\begin{equation*}
g_{\gamma(g)}^L=-c_{F1(G)}\frac{2v m_t}{m_\rho^2},\qquad g_{\gamma(g)}^R=-c_{F2(G)}\frac{2v m_t}{m_\rho^2}.
\end{equation*}
From the power counting of eq.\eqref{NDA} the size of the coefficients $c_I$ are easily estimated to be:
\begin{equation}\label{topFCNCcoupling}
c_{H1(2)}\sim g_\rho^2\epsilon_i^q\epsilon_3^q,\qquad c_{H3}\sim g_\rho^2\epsilon_i^u\epsilon_3^u,
\end{equation}
\begin{equation*}
c_{F1(G)}\sim g_\rho\left(\frac{g_\rho}{4\pi}\right)^2\epsilon_i^u\epsilon_3^q,\qquad c_{F2(G)}\sim g_\rho\left(\frac{g_\rho}{4\pi}\right)^2\epsilon_i^q\epsilon_3^u,
\end{equation*}

Using the results of Section~\ref{framework} it follows that the operators involving the charm quark are the most important ones. Among them, the chirality breaking operators $\bar c_R\Gamma t_L$ and $\bar c_L\Gamma t_R$ are equally relevant:
\begin{equation}
\frac{c_{F(G)1}}{c_{F(G)2}}\sim \frac{m_c}{m_t}\frac{1}{\lambda^4}\sim 1.5.
\end{equation}
Finally right-handed currents give the dominant contribution to $t\to qZ$ decays unless $\epsilon_3^u/\epsilon_3^q\lesssim 0.8$. 

From eq.~\ref{lagtopFCNC}, and neglecting the mass of the final state quarks, one finds the following FCNC width for the top quark \cite{Giudice:2012qq}:
\begin{eqnarray}\label{topwidth}
\Gamma(t\to q^iZ)&=& \frac{g^2}{128\pi \cos^2\theta_w}|g_Z^i|^2\frac{m_t^3}{m_Z^2}\left(1-\frac{m_Z^2}{m_t^2}\right)^2\left(1+2\frac{m_Z^2}{m_t^2}\right),\\\no
\Gamma(t\to q^i\gamma)&=&\frac{\alpha_{em}}{4}|g_\gamma^i|^2m_t,\\\no
\Gamma(t\to q^ig)&=&\frac{\alpha_{s}}{3}|g_g^i|^2m_t,
\end{eqnarray}
where $|g_X^i|^2\equiv |g_X^{Li}|^2+|g_X^{Ri}|^2$, $X=Z,\gamma,g$. As discussed in \cite{meletopfcnc}, the $t\to qg$ channel is swamped by the QCD background while sensitivities of the order $10^{-4}$ and $10^{-5}$ can be reached at the LHC for $BR(t\to q Z)$ and $BR(t\to q\gamma)$ respectively. Plugging eq.~\ref{topFCNCcoupling} in eq.~\ref{topwidth}, and taking $\Lambda=10$~TeV we find that the $t\to c\gamma$ branching ratio turns out to be negligibly small even in the most favorable scenario $g_\rho\sim 4\pi$. Focusing on the operator $\mathcal O^t_{H3}$ one obtains 
\begin{equation}
BR(t\to cZ)\sim 2\cdot 10^{-5}\left(\frac{10~{\rm TeV}}{\Lambda}\right)^4(\epsilon_3^u)^4.
\end{equation}
Only with a mild enhancement of the coefficient of $\mathcal O^t_{H3}$ above its natural value, and a mostly composite right-handed top, a measurable $t\to c Z$ rate will thus be detectable at the LHC.

\subsection{Lepton Flavor Violation}
\label{leptons}

While the EDM of the muon is safely below the present bounds for $\Lambda\sim10$ TeV, from Table \ref{table:generic} we see that the electron EDM is predicted to be $\sim10^2$ times bigger. Therefore, unless additional assumptions are made on how CP is broken, models of Partial Compositeness fail at describing flavor in the lepton sector.


For the case of $\mu\to e\gamma$ we write the effective Lagrangian as:
\ba
{\cal L}_{\mu\to e\gamma}=m_\mu e F_{\mu\nu}\left(\frac{\overline\mu_L\sigma^{\mu\nu} e_R}{\Lambda_L^2}+\frac{\overline\mu_R\sigma^{\mu\nu} e_L}{\Lambda_R^2}\right),
\ea
from which we derive:
\ba\label{BR}
{\rm BR}(\mu\to e \gamma)=384\,\pi^2 e^2\left(\left|\frac{v}{\Lambda_L}\right|^4+\left|\frac{v}{\Lambda_R}\right|^4\right).
\ea

Requiring BR$(\mu\to e \gamma)<2.4\times10^{-12}$~\cite{Adam:2011ch} gives $\Lambda_{L,R}\gtrsim600$ TeV. Using~(\ref{NDA}) we write:
\ba\label{learn}
\frac{m_\mu}{\Lambda_L^2}&=&\frac{m_e}{\Lambda^2}c_{21}^{\ell e}\frac{\epsilon_2^\ell}{\epsilon_1^\ell}=\frac{\sqrt{m_em_\mu}}{\Lambda^2}c_{21}^{\ell e}\left(\frac{\epsilon_1^\ell}{\epsilon_2^\ell}\sqrt{\frac{m_\mu}{m_e}}\right)^{-1},\\\no
\frac{m_\mu}{\Lambda_R^2}&=&\frac{m_\mu}{\Lambda^2}c_{21}^{e\ell}\frac{\epsilon_1^\ell}{\epsilon_2^\ell}=\frac{\sqrt{m_em_\mu}}{\Lambda^2}c_{21}^{e\ell}\left(\frac{\epsilon_1^\ell}{\epsilon_2^\ell}\sqrt{\frac{m_\mu}{m_e}}\right).
\ea
Eq.~(\ref{learn}) has been written so as to emphasize that the optimal choice for the mixing parameters in the lepton sector is shown in Eq.(\ref{R}). Imposing this ansatz, we can finally translate the bound into constraints on the $O(1)$ coefficients $c_{21}^{\ell e}$, $c_{21}^{e\ell}$ as shown in Table~\ref{table:generic}. 

For the muon conversion inside nuclei the relevant Lagrangian is of the form~(\ref{penguin}). We follow the notation of~\cite{lfv} and write the ratio between conversion and capture rates for a generic nuclei $N$ as:
\begin{equation}
B_{\mu\to e}^{N}=\left|\frac{(4\pi)^2}{\Lambda^2}(\epsilon_1^\ell\epsilon_2^\ell~c_{12}^{\ell\ell}+\epsilon_1^e\epsilon_2^e~c_{12}^{ee})\right|^2\left[\left(1-4s_W^2\right) V_N^{(p)}-V_N^{(n)}\right]^2\frac{m_\mu^5}{\Gamma^{N}_{\rm capt.}},
\end{equation}
where $\Gamma^{N}_{\rm capt.}$ is the capture rate while $V_N^{(p,n)}$ are nuclear form factors.
Currently, the strongest bounds on $B_{\mu\to e}^{N}$ are from targets of gold and titanium~\cite{PDG}. 
We find that the most stringent constraint arises from $\mu\to e$ conversions in gold, with $B_{\mu\to e}^{Au}<7\times10^{-13}$.
The reason is that the larger atomic weight of $Au$ leads to a much more effective coherent conversion, thus resulting in bigger form factors. 
Using~\cite{lfv}:
\begin{equation*}
V_{Au}^{(p)}\approx0.09\quad~~~~~~~~~~~~~~~~V_{Au}^{(n)}\approx0.1
\end{equation*}
and $\Gamma^{Au}_{\rm capt.}= 8.7\times 10^{-18}$ GeV, and assuming that a single operator at a time is switched on, the experimental bound becomes:
\begin{equation}
c_{12}^{\ell\ell}\lesssim5\times10^{-6}\frac{1}{\epsilon_1^\ell \epsilon_2^\ell}
,~~~~~~~~~c_{12}^{ee}\lesssim5\times10^{-6}\frac{1}{\epsilon_1^e \epsilon_2^e}.
\end{equation}
The bound is roughly an order of magnitude stronger than that from $\mu\to e\overline{e}e$ transitions, also mediated by~(\ref{eq:gen:1bis}).

\subsection{Electro-Weak Precision Tests}

The dominant corrections to the oblique electroweak precision observables can be encoded in the operators
\begin{equation} \label{eq:ewpt}
{\cal L}_{\rm EWPT} \supset c_S\frac{gg'}{m_\rho^2}H^\dagger W_{\mu\nu}HB^{\mu\nu}+c_T\frac{g_\rho^2}{m_\rho^2}|H^\dagger D_\mu H|^2
\end{equation}
from which we get:
\ba\label{generic}
\widehat S&=&c_S\frac{m_W^2}{m_\rho^2}=6.4\times10^{-5}c_S\left(\frac{10~{\rm TeV}}{\Lambda}\right)^2\left(\frac{4\pi}{g_\rho}\right)^2\\\no
\widehat T&=&-c_T\frac{g_\rho^2v^2}{m_\rho^2}=-4.8\times10^{-2}c_T\left(\frac{10~{\rm TeV}}{\Lambda}\right)^2.
\ea
For $c_{S}\sim1$ and $\Lambda\sim10$ TeV we see that $\widehat S$ is well within the experimental bound $|\widehat S|\lesssim 2\cdot10^{-3}$ for a moderately large $g_\rho$. Additional, IR contributions from Higgs exchange are also under control. Despite the large value of $\Lambda$, the $\widehat T$ parameter exceeds the bound by a factor $O(10)$ if $c_T\sim1$. One can suppress $c_T$ assuming that the composite sector respects the full $SU(2)_L\times SU(2)_R$ custodial symmetry. In this case the dominant contributions to $\widehat T$ are determined by the couplings of the top to the composite sector 
\begin{equation}
\mathcal L\supset \epsilon_3^q\,\bar q_{3L} O_R+ \epsilon_3^u\,\bar u_{3R} O_L,
\end{equation}
and by the observation that $\widehat T$ transforms as a $\mathbf 5$ of custodial isospin. Since the strong sector respects custodial symmetry we can assign the $\epsilon$s spurious $SU(2)_L\times SU(2)_R$ quantum numbers in order to understand their contributions to $\widehat T$. Working under the assumption of a single source of custodial breaking the simplest possibilities are (in the following the index in the representation is the $U(1)_X$ quantum number needed to couple the fermions and which is linked to the hypercharge by $Y=T_{3R}+X$):
\begin{itemize}
\item $O_L=(\mathbf 2, \mathbf 1)_{1/6}$,  $O_R=(\mathbf 1, \mathbf 2)_{1/6}$, $c_T=(\epsilon_3^u)^4\left(\frac{g_\rho}{4\pi}\right)^2$,
\item $O_L=(\mathbf 2, \mathbf 2)_{2/3}$,  $O_R=(\mathbf 1, \mathbf 1)_{2/3}$, $c_T=(\epsilon_3^q)^4\left(\frac{g_\rho}{4\pi}\right)^2$.
\end{itemize}
In the strong limit $g_\rho\sim 4\pi$, $\epsilon_3^{u(q)}$ has to be smaller than $\sim 0.4$.
If we allow one of $O_{L,R}$ to be a triplet under $SU(2)_R$ then only two powers of $\epsilon_3^{q,u}$ are needed to saturate the quantum numbers of $\widehat T$ and the bound is stronger, $\epsilon_3^{u(q)}\lesssim 0.2$.

Other constraints arise from the following set of operators which modify the $Zb_Lb_L$ coupling:
\begin{equation} \label{eq:ewpt}
{\cal L}_{\rm EWPT} \supset  \frac{g_\rho^2 (\epsilon_3^q)^2}{m_\rho^2} \left( c_{H1}\bar q_{3L}\gamma^\mu q_{3L}\, iH^\dagger  \overleftrightarrow D_\mu H+c_{H2}\bar q_{3L}\gamma^\mu \tau^aq_{3L}\, i H^\dagger \tau^a  \overleftrightarrow D_\mu H \right) \, .
\end{equation}
Such operators can be generated after EWSB if $O_{L(R)}$ interpolates for charge $-1/3$ states which can mix with the bottom quark. Focusing on the first one gets
\ba\label{Zbbar}
\left|\frac{\delta g_{b_L}}{g_{b_L}}\right|&=&2 |c_{H1}|(\epsilon_3^q)^2(4\pi)^2\frac{v^2}{\Lambda^2}
= 9.6\times10^{-2}|c_{H1}|(\epsilon_3^q)^2\left(\frac{10~{\rm TeV}}{\Lambda}\right)^2,
\ea
with the experimental limit from \cite{ALEPH:2005ab} reading $\left|\frac{\delta g_{b_L}}{g_{b_L}}\right|\lesssim0.25\%$. Applying this to eq. \eqref{Zbbar} gives $\epsilon_3^q\lesssim 0.15$. This is fine if $O_L=(\mathbf 2, \mathbf 1)$ but gives a tension with the top quark mass if $O_L=(\mathbf 2, \mathbf 2)$, and even more so if the left-handed top mixes with an $SU(2)_R$ triplet. It has been suggested in ~\cite{Agashe:2006at} that, assuming the strong sector to be symmetric under the full $O(4)\supset SU(2)_L\times SU(2)_R$ and embedding $O_L$ in appropriate representations of the custodial group, one can use the extra parity in $O(4)$ to impose $\delta g_{b_L}=0$. This $Z_2$ symmetry has been shown in~\cite{ratt2HDM} to be a generic accidental symmetry, at the level of the 2-derivative $\sigma$-model, in many of the pNGB Higgs constructions in the literature.

\section{Phenomenological Constraints in the SUSY case}
\label{SUSYconstraints}

Using (\ref{eq:deltas:susy}) and taking as input parameters those in (\ref{eq:susy:assumptions}) we can repeat the analysis of the flavor bounds of Section \ref{sec:general}, setting upper limits on the coefficients $(c^{f}_{ij})_{LR, RL, LL, RR}$. 
Let us go briefly through the derivation of the more stringent bounds of Table \ref{table:susy}.

\subsection{Dipole Operators in the Quark Sector}

The analysis of the phenomenological bounds proceed in complete analogy with the generic case of section~\ref{pheno}, except for the fact that now the same mass insertion can contribute to different Wilson coefficients of the low energy theory, and possible interference effects must be taken into account.

The strongest constraint on the present framework arises from the neutron EDM. At the scale $\tilde m$ we have, according to~\cite{Gabbiani:1996hi}:
\ba
\frac{d_d(\tilde m)}{e}&=&-\frac{2}{9}\frac{\alpha_s}{\pi}\frac{m_{\tilde g}}{m_{\tilde q}^2}M_1(x){\rm Im}(\delta_{11}^d)_{LR}\\\no
\frac{d_u(\tilde m)}{e}&=&\frac{4}{9}\frac{\alpha_s}{\pi}\frac{m_{\tilde g}}{m_{\tilde q}^2}M_1(x){\rm Im}(\delta_{11}^u)_{LR}\\\no
\frac{\tilde d_q(\tilde m)}{e}&=&\frac{1}{4}\frac{\alpha_s}{\pi}\frac{m_{\tilde g}}{m_{\tilde q}^2}\left(-\frac{1}{3}M_1(x)-3M_2(x)\right){\rm Im}(\delta_{11}^q)_{LR}~~~~~~~~~(q=u,d).
\ea
Here $x=m_{\tilde g}^2/m_{\tilde q}^2$ and the loop functions $M_i(x)$ are defined in \cite{Gabbiani:1996hi}. The leading effect of the running from the scale $\tilde m$ to $\mu\sim1$ GeV is given by (see for instance~\cite{Buras:1999da}):
\ba
d_q(\mu)&=&\eta^{4/3\beta_0}d_q(\tilde m)+8\left(\eta^{4/3\beta_0}-\eta^{2/3\beta_0}\right)eQ_q\tilde d_q(\tilde m)\\\no
\tilde d_q(\mu)&=&\eta^{2/3\beta_0}\tilde d_q(\tilde m),
\ea
where $\eta=[\alpha_s(\mu)/\alpha_s(\tilde m)]$ and $Q_{u(d)}=2/3(-1/3)$. Imposing the bound~\cite{PDG}, and assuming that only one $(\delta_{11}^{q})_{LR}$ at a time is turned on, we find the results shown in Table \ref{table:susy}.
There are no significant cancellations among the various terms in Eq.~(\ref{nEDM}) because the chromoelectric contributions tend to dominate.

Similarly to what happens in the non-SUSY scenario, $\epsilon'/\epsilon$ provides another important constraint. Using the same notation of Appendix~\ref{pheno}, we write~\cite{Gabbiani:1996hi}:
\ba\no
C_G&=&\frac{m_s}{v}\frac{\alpha_s}{8\pi}\frac{1}{m_{\tilde q}^2}\left[\left(-\frac{1}{3}M_3(x)-3M_4(x)\right)(\delta_{12}^d)_{LL}+\frac{m_{\tilde g}}{m_s}\left(-\frac{1}{3}M_1(x)-3M_2(x)\right)(\delta_{12}^d)_{LR}\right]\\\no
C_G'&=&\frac{m_s}{v}\frac{\alpha_s}{8\pi}\frac{1}{m_{\tilde q}^2}\left[\left(-\frac{1}{3}M_3(x)-3M_4(x)\right)(\delta_{12}^d)_{RR}+\frac{m_{\tilde g}}{m_s}\left(-\frac{1}{3}M_1(x)-3M_2(x)\right)(\delta_{12}^d)_{RL}\right].
\ea
Again, the above Wilson coefficients are evaluated at the soft SUSY breaking scale. The bounds shown in Table \ref{table:susy} are derived imposing condition~(\ref{run}) and assuming maximal CP-violating phases, as usual. In particular, we find the relatively severe bound Im$(\delta_{12}^d)_{LR,RL}\lesssim3-4\times10^{-5}$ for $m_{\tilde g}\sim m_{\tilde q}\sim1$ TeV.

\subsection{Dipole Operators in the Lepton Sector}

Let us now discuss in some detail the most stringent observable processes in the lepton sector. These are $\mu\rightarrow e\gamma$ and the electron EDM. For definiteness we work under the assumption of a degenerate spectrum of sleptons with masses $\tilde m$.

Using the results of \cite{Rattazzi:1995ts}\cite{Masina:2002mv}\cite{Paradisi:2005fk}, we can write:
\begin{eqnarray}
&&\frac{BR(\mu \rightarrow e\gamma)}{BR(\mu \rightarrow e \overline{\nu}_e \nu_{\mu})} \quad = \quad
\frac{48 \pi^3 \alpha_{em}}{G_F^2} \times \left\{
\left| \frac{g^{\prime \, 2}}{16 \pi^2}  \, \frac{ (\delta^{e}_{12})_{RL}  \, M_{\tilde{B}}}{m_{\mu} \tilde{m}^2} f(\frac{M_{\tilde{B}}^2}{\tilde{m}^2})  \right|^2 \right. + \nonumber\\
&&\left. + \left| \frac{g^{\prime \, 2}}{16 \pi^2}  \, \frac{  (\delta^{e}_{12})_{LR} \, M_{\tilde{B}}}{m_{\mu} \tilde{m}^2} f(\frac{M_{\tilde{B}}^2}{\tilde{m}^2})
+
 \frac{g^2}{16 \pi^2}  \, \frac{ (\delta^{e}_{12})_{LL} \, M_{\tilde{W}}\mu \tan\beta}{(M_{\tilde{W}}^2 - \mu^2) \tilde{m}^2} \left( g(\frac{M_{\tilde{W}}^2}{\tilde{m}^2}) - g(\frac{\mu^2}{\tilde{m}^2}) \right) +
\right.\right. \nonumber \\
&& \left.\left. +
\frac{g^{\prime \, 2}}{16 \pi^2}  \, \frac{ (\delta^{e}_{12})_{LL}\, M_{\tilde{B}}\mu \tan\beta }{ \tilde{m}^2}   \left( -\frac{  f(\frac{M_{\tilde{B}}^2}{\tilde{m}^2})-f(\frac{\mu^2}{\tilde{m}^2}) }{2(M_{\tilde{B}}^2 - \mu^2)}  
+ \frac{h(\frac{M_{\tilde{B}}^2}{\tilde{m}^2})}{\tilde{m}^2} \right)
\right|^2 \right\}.
\end{eqnarray}
In the above expression we neglected the subdominant contribution from $(\delta^{e}_{12})_{RR}$, which receives contributions only from the bino loop, while in the Left-Left insertions we only kept the terms that are $\tan\beta-$enhanced. The loop functions are defined as:
\begin{eqnarray}
& f(x) = \frac{1+4x-5x^2 +2x(2+x)\log x}{2(1-x)^4}
\\
& g(x) = \frac{11-4x-7x^2 +2(2+6x+x^2)\log x}{4(1-x)^4}
\\
& h(x) = \frac{-1-9x+9x^2+x^3-6x(1+x)\log x}{2(x-1)^5},
\end{eqnarray}
whereas $M_{\tilde{B}}$ and $M_{\tilde{W}}$ are the bino and wino masses, respectively. Note that only the bino contributes to the diagram involving the LR and RL mass insertions.

To estimate the rate we take~\footnote{The assumption $2\mu\sim M_{\tilde{W}}$ is conservative, since for $\mu\sim M_{\tilde{W}}$ the Wino contribution tends to be suppressed.} $M_{\tilde{B}}\sim M_{\tilde{W}}\sim 2 \mu \sim \tilde{m} \sim $ 1 TeV.
Switching on only one $\delta_{12}^{e}$ at a time and imposing the experimental bound of \cite{Adam:2011ch} we obtain:
\begin{equation}
(\delta_{12}^{e})_{LR,RL}< 8 \times 10^{-6},
\quad\quad\quad\quad
(\delta_{12}^{e})_{LL} < \frac{1 \times 10^{-2}}{\tan\beta} \, .
\end{equation}
Finally, one can derive the bounds on the dimensionless coefficients defined in~(\ref{massinse}). The results are shown in Table \ref{table:susy} for the optimal choice~(\ref{R}).

In the case of the electron EDM, we consider the bound on the flavor-blind part coming from the $A-$term, i.e. the bound on $(\delta^{e}_{11})_{LR}$, since the contributions involving more mass insertions and/or insertions of the electron mass are subdominant. Using the results of \cite{Masina:2002mv} we can write,  in the limit of degenerate slepton masses:
\begin{equation}
\frac{d_e}{e} = -\frac{g^{\prime \,2}}{16 \pi^2} \, \frac{ \, \mbox{Im}\left[ M_{\tilde{B}} (\delta^{e}_{11})_{LR} \right]  }{ \tilde{m}^2 } \, f(\frac{M_{\tilde{B}}^2}{\tilde{m}^2}) \, .
\end{equation}
Imposing the experimental bound $|d_e| < 1.6 \times 10^{-27} \, e$\, cm \cite{Regan:2002ta} and conservatively assuming  a phase of order 1, we obtain for $M_{\tilde{B}} \sim \tilde{m}$:
\begin{equation}
(\delta_{11}^{e})_{LR}< 7 \times 10^{-7} \, ,
\end{equation}
from which one obtains the bound reported in the Table.
Notice that there are also other flavor-blind contributions enhanced by $\tan\beta$ coming from the imaginary part of the combination gaugino masses $\times$ $\mu$ term. Although these terms can be important and give effects at the level of the experimental sensitivity, we do not consider them here since they do not give any constraint on our mass insertions as defined in (\ref{eq:deltas:susy}).
They do give a bound on the phase of the $\mu$ term times gaugino masses which is of the same order of the bound on $(c_{11}^{e})_{LR}$, with more dependence on the precise value of $|\mu|$.


\vspace{0.3cm}

\begin{multicols}{2}

\end{multicols}

\end{document}